\documentclass[sigconf]{acmart}

\usepackage{enumitem}
\usepackage{makecell}
\usepackage{multirow}
\usepackage{fontawesome}
\usepackage{subcaption}
\usepackage{verbatim}
\usepackage{pifont}
\usepackage{rotating}
\usepackage{soul}
\usepackage[lined,boxed,linesnumbered,commentsnumbered,ruled]{algorithm2e}
\usepackage{color}
\usepackage{pifont}
\usepackage{colortbl}
\usepackage{framed}
\usepackage{xcolor}
\definecolor{mygray}{gray}{.9}
\definecolor{lightbluebg}{RGB}{245,248,255}

\newenvironment{mybreakablebox}{
    
    \MakeFramed{\advance\hsize-\width \FrameRestore}
    \scriptsize
}{
    \endMakeFramed
}

\newcommand{\tech}{\textit{REAgent}}

\newcommand{\Comment}[1]{}

\NewDocumentCommand{\framecolorbox}{oommm}
 {
  \IfValueTF{#1}
   {\IfValueTF{#2}
    {\fcolorbox{#3}{#4}{\makebox[#1][#2]{#5}}}
    {\fcolorbox{#3}{#4}{\makebox[#1]{#5}}}%
   }
   {\fcolorbox{#3}{#4}{#5}}%
 }

\AtBeginDocument{
  \providecommand\BibTeX{{
    \normalfont B\kern-0.5em{\scshape i\kern-0.25em b}\kern-0.8em\TeX}}}

\renewcommand\footnotetextcopyrightpermission[1]{} 

\usepackage[most]{tcolorbox}
\tcbset{
  colback=blue!4,
  boxrule=0pt,
  frame hidden,
  arc=3pt,
  left=6pt,
  right=6pt,
  top=6pt,
  bottom=6pt,
  fontupper=\scriptsize
}

\begin{document}

\title{REAgent: Requirement-Driven LLM Agents for Software Issue Resolution}

\author{Shiqi Kuang}
\orcid{0009-0008-0532-6655}
\affiliation{
  \institution{School of Computer Software, Tianjin University}
  \city{Tianjin}
  \country{China}
}
\email{kuangshiqi@tju.edu.cn}

\author{Zhao Tian}
\orcid{0000-0002-9316-7250}
\affiliation{
  \institution{School of Computer Software, Tianjin University}
  \city{Tianjin}
  \country{China}
}
\email{tianzhao@tju.edu.cn}

\author{Kaiwei Lin}
\orcid{}
\affiliation{
  \institution{The International Joint Institute of Tianjin University, Tianjin University}
  \city{Tianjin}
  \country{China}
}
\email{linkaiwei_1012@tju.edu.cn}

\author{Chaofan Tao}
\orcid{}
\affiliation{
  \institution{HUAWEI Technologies}
  \country{China}
}
\email{tao.chaofan@huawei.com}

\author{Shaowei Wang}
\orcid{}
\affiliation{
  \institution{HUAWEI Technologies}
  \country{China}
}
\email{wangshaowei19@huawei.com}

\author{Haoli Bai}
\orcid{}
\affiliation{
  \institution{HUAWEI Technologies}
  \country{China}
}
\email{baihaoli@huawei.com}

\author{Lifeng Shang}
\orcid{}
\affiliation{
  \institution{HUAWEI Technologies}
  \country{China}
}
\email{Shang.Lifeng@huawei.com}

\author{Junjie Chen}
\authornote{Junjie Chen is the corresponding author.}
\orcid{0000-0003-3056-9962}
\affiliation{
  \institution{School of Computer Software, Tianjin University}
  \city{Tianjin}
  \country{China}
}
\email{junjiechen@tju.edu.cn}

\begin{abstract}
Issue resolution aims to automatically generate patches from given issue descriptions and has attracted significant attention with the rapid advancement of large language models (LLMs).
However, due to the complexity of software issues and codebases, LLM-generated patches often fail to resolve corresponding issues. 
Although various advanced techniques have been proposed with carefully designed tools and workflows, they typically treat issue descriptions as direct inputs and largely overlook their quality (e.g., missing critical context or containing ambiguous information), which hinders LLMs from accurate understanding and resolution.
To address this limitation, we draw on principles from software requirements engineering and propose \tech{}, a requirement-driven LLM agent framework that introduces \textit{issue-oriented requirements} as structured task specifications to better guide patch generation. 
Specifically, \tech{} automatically constructs structured and information-rich issue-oriented requirements, identifies low-quality requirements, and iteratively refines them to improve patch correctness.
We conduct comprehensive experiments on three widely used benchmarks using two advanced LLMs, comparing against five representative or state-of-the-art baselines. 
The results demonstrate that \tech{} consistently outperforms all baselines, achieving an average improvement of 17.40\% in terms of the number of successfully-resolved issues (\textit{\% Resolved}).
\end{abstract}



\maketitle
\section{Introduction} 
\label{sec:intro}
Issue resolution aims to automatically generate code patches that satisfy requirements described in software repository issues, thereby fixing defects or implementing feature requests~\cite{jimenez2024swe, jiang2025agentic, zhang2023survey, tao2024magis}. 
Effective issue resolution techniques can substantially improve developer productivity~\cite{peng2023impact, tao2024magis}, enhance software quality~\cite{liu2023your, xia2023automated}, and reduce the manual effort required for localization and repair~\cite{tao2024magis, xia2025demystifying}. 
Recent advances in large language models (LLMs), such as DeepSeek~\cite{guo2024deepseek} and Qwen~\cite{cao2026qwen3}, have led to significant progress in code-related tasks. 
These LLMs demonstrate strong capabilities in code generation and understanding, and are increasingly applied in software engineering scenarios~\cite{kuang2025effectiveness, shrivastava2023repofusion, jiang2025agentic, jiang2026survey}. 
Despite their success on function-level tasks, LLMs still struggle with repository-level issue resolution~\cite{meng2024empirical, aleithan2024swe, deng2025swe}. 
For example, DeepSeek-V3.2 achieves 83.30\% accuracy on the function-level benchmark LiveCodeBench~\cite{jainlivecodebench}, but only 15.56\% on the repository-level benchmark SWE-bench Pro~\cite{deng2025swe}. 
This significant performance gap highlights the fundamental challenges of resolving complex repository-level issues.

To bridge this gap, prior work has proposed agent- and workflow-based techniques that enhance LLMs with tool use and structured workflows. 
They enable models to iteratively explore repositories, retrieve relevant code, and validate generated patches. 
For example, SWE-agent~\cite{yang2024swe} equips LLMs with tools such as file retrieval, code search, and test execution to facilitate repository interaction. 
Agentless~\cite{xia2025demystifying} decomposes issue resolution into predefined stages, including localization, patch generation, and patch validation. 
Subsequent work further improves these frameworks by introducing advanced retrieval strategies~\cite{ouyang2025repograph, chen2025prometheus}, context compression methods~\cite{wang2026swe, lindenbauer2025complexity}, and multi-agent collaboration mechanisms~\cite{chen2024coder, pabba2025semagent}.

Despite these advances, existing techniques primarily focus on improving how LLMs solve problems through better tools or workflows, while largely overlooking what is being solved, namely the quality of the task specification itself. 
Most techniques directly treat issue descriptions as input, implicitly assuming that they accurately capture the programming specifications for the desired code patches. 
In practice, however, this assumption rarely holds.
Specifically, issue descriptions are written in natural language by users or developers to document system anomalies or feature requests. 
Their main purpose is to facilitate human communication, not to serve as precise implementation specifications for patch generation. 
Consequently, they often lack critical contextual information and contain ambiguous or incomplete descriptions~\cite{yang2023users, bettenburg2008makes, zimmermann2010makes, chaparro2017detecting, davies2014s, suri2026codescout}. 
For example, more than 70\% of issues lack essential elements such as reproduction steps or validation criteria~\cite{soltani2020significance}, making them difficult to interpret and resolve. 
Moreover, the use of unstructured natural language introduces subjectivity, leading to inconsistent interpretations across developers~\cite{huang2019empirical}. 
These limitations fundamentally hinder LLMs, which are highly sensitive to input quality, from accurately understanding and resolving issues.


\textit{This observation suggests that the bottleneck of repository-level issue resolution lies not only in model capability or reasoning strategy, but also in the lack of high-quality task specification.}
Insights from software requirements engineering highlight the value of structured artifacts in systematically capturing system behavior and constraints~\cite{ouhbi2013software, van2008requirements}. 
Such artifacts typically organize information into key elements, including background context, functional goals, system environment, behavioral constraints, and verifiable success criteria~\cite{montgomery2022empirical, stephen2020evaluation}. 
Inspired by this principle, we argue that constructing structured requirements for patch generation can substantially improve issue resolution.
In the context of issue resolution, where both the issue and the codebase are already available, we refer to such structured representations as \textbf{issue-oriented requirements} to distinguish them from traditional software requirements often defined prior to development. 
By analogy, incorporating structured elements into issue-oriented requirements enables the supplementation of missing contextual information in issues and reduces ambiguity by explicitly defining task objectives, modification scope, and constraints. 
That is, constructing structured and information-rich issue-oriented requirements from issues and repository context can provide LLM-based agents with clearer and more precise guidance, thereby representing a promising direction for improving the effectiveness of repository-level issue resolution.


However, driving issue resolution through issue-oriented requirements still faces three key challenges.
(1) \textbf{Difficulty in collecting and organizing scattered information}. 
Issue-oriented requirements must capture extensive, issue-specific information scattered across multiple files and modules in the repository. 
Accurately and efficiently retrieving and integrating such information from large codebases, and organizing it into a structured representation that effectively guides patch generation, is inherently difficult.
(2) \textbf{Difficulty in evaluating requirement quality}. 
Due to the complexity of real-world tasks and the inherent hallucination tendencies of LLMs, generating high-quality issue-oriented requirements in a single attempt remains highly challenging. 
Low-quality requirements can, in turn, adversely affect the correctness of subsequently generated patches. 
Therefore, accurately assessing requirement quality is quite necessary.
However, requirements are expressed in structured natural language and exhibit a degree of undecidability~\cite{ferrari2014pragmatic}, making their quality difficult to evaluate using simple rules or static analysis methods~\cite{gervasi2002lightweight}.
(3) \textbf{Difficulty in fixing requirement deficiencies}. 
Even when requirement quality deficiencies are identified, effectively correcting them remains challenging. 
The large semantic space (characterized by complex requirement attributes and extensive requirement expressions) makes it difficult to pinpoint root causes.
Meanwhile, the lack of actionable feedback (i.e., root causes and corresponding refinement guidelines) further hinders targeted and efficient refinement. 

To address these challenges, we propose \textbf{\tech{}}, a novel requirement-driven LLM agent approach for repository-level issue resolution. 
Specifically, \tech{} automatically generates structured and information-rich issue-oriented requirements, identifies low-quality requirements, and iteratively refines requirements, facilitating more effective patch generation for issue resolution.
To address the first challenge, we design a \textbf{requirement generation} component. 
It employs a requirement generation agent that autonomously explores the complex codebase to collect issue-specific contextual information and systematically applies a series of pre-defined requirement attributes to construct structured and information-rich requirements.
To address the second challenge, we design a \textbf{requirement assessment} component. 
It introduces a requirement assessment agent that leverages traceability between requirements and generated patches~\cite{mucha2024systematic, yoo2024building} to transform requirement evaluation into patch assessment. 
Using executable results as indirect quality signals, we define the Requirement Assessment Score (RAS) to measure how well the requirements guide correct implementations. 
To address the third challenge, we design a \textbf{requirement refinement} component. 
It categorizes root causes of low-quality requirements into three high-level classes of requirement deficiencies and develops tailored refinement strategies for each category, thereby reducing the space of requirement refinement.
Specifically, we design a requirement analysis agent, which first determines the deficiency category for a given low-quality requirement and then applies category-specific strategies to generate actionable feedback that effectively guides requirement refinement.

Based on three widely used repository-level issue resolution benchmarks (i.e., SWE-bench Lite~\cite{jimenez2024swe}, SWE-bench Verified~\cite{openai_swebench_verified}, SWE-bench Pro~\cite{deng2025swe}), we conduct a comprehensive evaluation of \tech{} on two advanced LLMs (i.e., DeepSeek-V3.2~\cite{liu2025deepseek} and Qwen-Plus~\cite{yang2025qwen3, bai2023qwen}). 
The results show that across all 6 experimental settings (2 LLMs $\times$ 3 benchmarks), \tech{} consistently outperforms 5 representative or state-of-the-art baselines. 
Specifically, compared with baselines, the number of instances successfully resolved by \tech{} increases by 9.17\%$\sim$24.83\% (i.e., \textit{\% Resolved}), and the number of instances with patches successfully applied increases by 22.17\%$\sim$49.50\% (i.e., \textit{\% Applied}). 
Then, we investigate the impact of the number of iterations ($N$), a critical hyper-parameter in all issue resolution techniques with iterative strategies.
The results indicate that as $N$ increases, \tech{} consistently outperforms all iterative baselines. 
Finally, we construct four variants of \tech{} for ablation studies, confirming the contribution of each main component.


\begin{figure}[t!]
    \centering
    \includegraphics[width=\columnwidth]{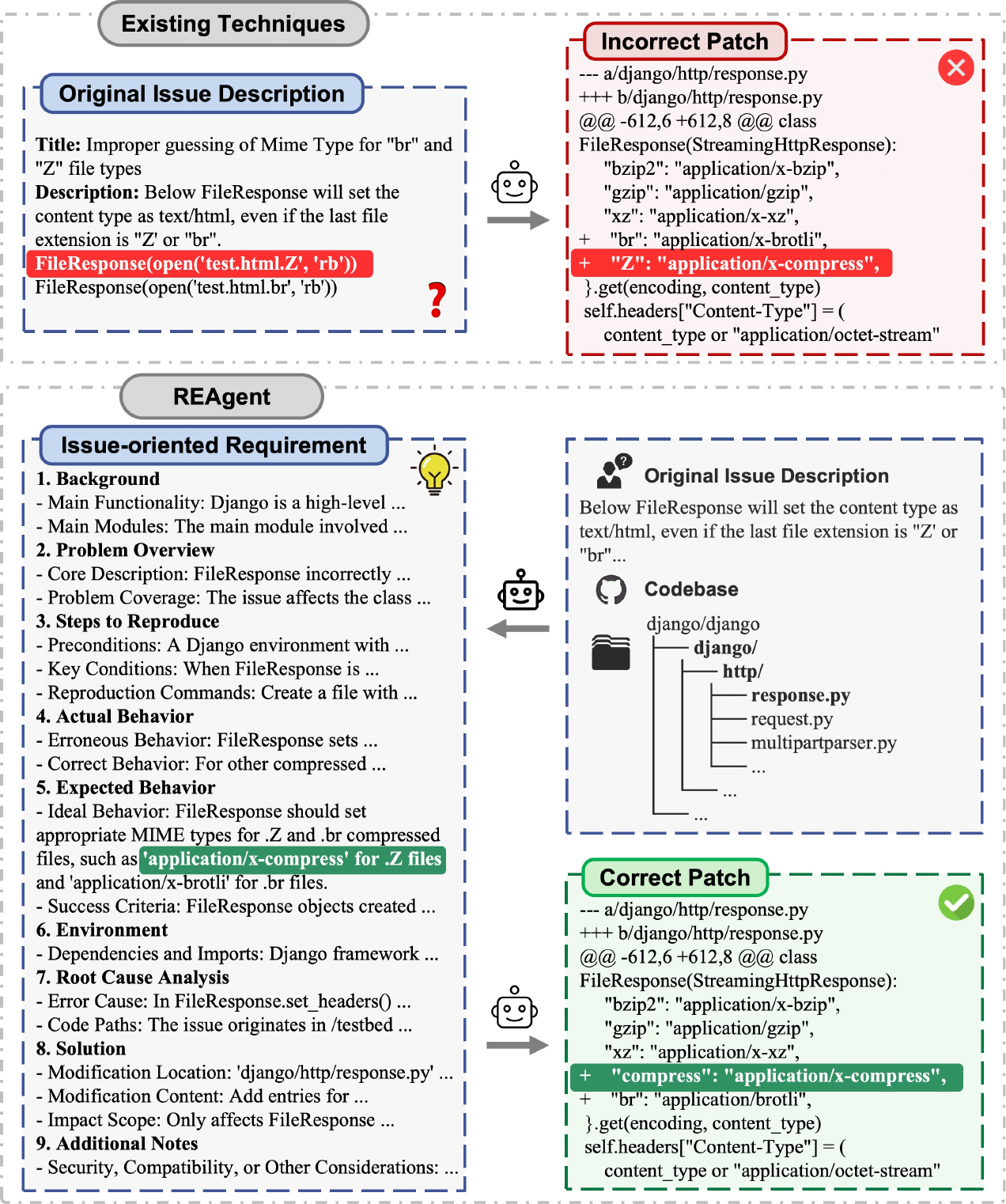}
    \vspace{-5mm}
    \caption{A real-world example from SWE-bench Verified with DeepSeek-V3.2}
    \label{fig:example}
    \vspace{-5mm}
\end{figure}

The main contributions of this paper are summarized as follows:
\begin{itemize}[noitemsep, topsep=0pt, leftmargin=*, align=left]
    \item \textbf{Novel Perspective:}
    We identify the quality of task inputs as a key bottleneck in repository-level issue resolution and introduce \textit{issue-oriented requirements} as structured task specifications to more effectively guide patch generation.
    

    \item \textbf{Requirement-Driven Framework:} We propose \tech{}, a requirement-driven LLM agent framework that systematically improves issue resolution through three components: (1) requirement generation, which constructs structured specifications from issues and codebases; (2) requirement assessment, which evaluates requirement quality using Requirement Assessment Score; and (3) requirement refinement, which identifies and corrects requirement deficiencies through targeted, iterative feedback.

    \item \textbf{Comprehensive Evaluation:} We conduct extensive experiments on three widely used benchmarks with two advanced LLMs, comparing against five representative or state-of-the-art baselines. Results show that \tech{} consistently achieves substantial improvements, demonstrating its effectiveness for repository-level issue resolution.


    
    
\end{itemize}
\section{Motivating Example}
\label{sec:example}

To illustrate the importance of issue-oriented requirements, we present a real-world case demonstrating the motivation of \tech{}. 
Figure~\ref{fig:example} shows an example from the SWE-bench Verified~\cite{openai_swebench_verified} dataset with instance\_id \textit{django\_\_django-16642}. 
We first employ an advanced LLM (DeepSeek-V3.2~\cite{guo2024deepseek}) within the state-of-the-art Trae-agent~\cite{gao2025trae} framework to generate a patch directly from the original issue description. 
However, due to the incompleteness and ambiguity of the issue description, the generated patch is incorrect. 
Specifically, the issue description fails to specify the encoding associated with the ``\texttt{.Z}'' file in \texttt{mimetypes.guess\_type()}, leading the agent to incorrectly assume that the encoding name is ``\texttt{Z}'', which results in an erroneous patch implementation.

In contrast, we employ \tech{} to solve the same issue using the same base model. 
Specifically, \tech{} first constructs a structured, issue-oriented requirement by leveraging both the original issue description and the codebase.
The resulting requirement explicitly captures key technical details, including \texttt{```application/x-compress' for .Z files''}. 
By supplementing the missing key information and resolving ambiguity in the original issue description, the issue-oriented requirement effectively guides the model to generate a correct patch.
This case study highlights the critical role of structured and information-rich issue-oriented requirements in improving the LLM performance in repository-level issue resolution.

\section{Approach}
\label{sec:approach}

\begin{figure*}[t]
    \centering
    \includegraphics[width=\textwidth]{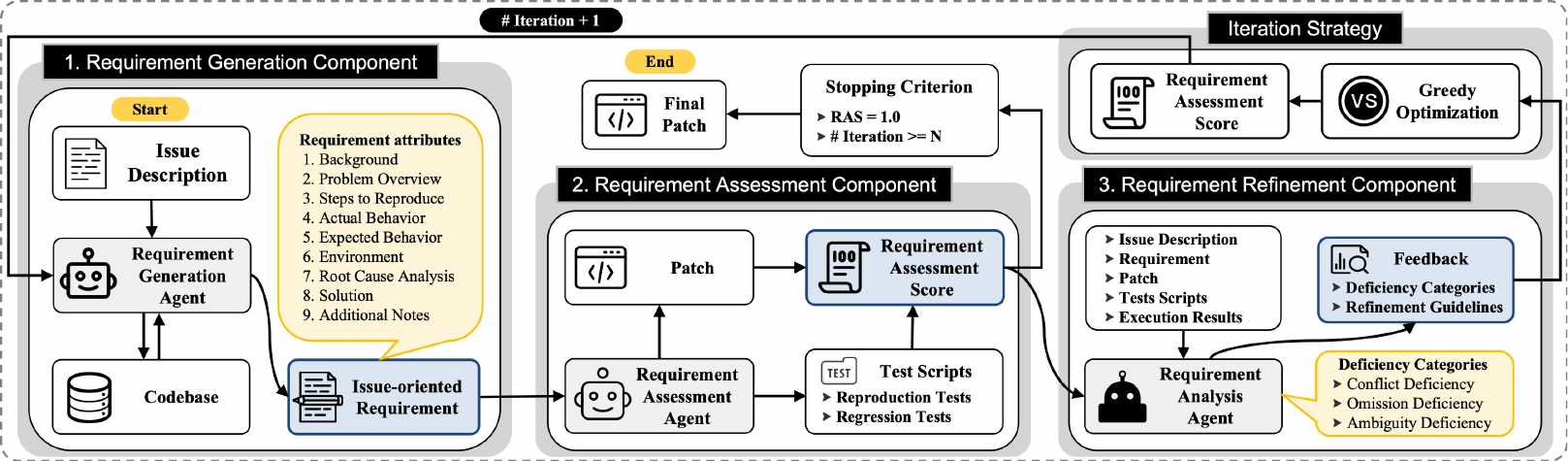}
    \vspace{-5mm}
    \caption{The overview of \tech{}}
    \vspace{-5mm}
    \label{fig:overview}
\end{figure*}


In this paper, we propose a novel requirement-driven LLM agent approach, \tech{}, to enhance the performance of repository-level issue resolution. 
It takes a simple issue description as input and automatically generates structured and information-rich issue-oriented requirements, identifies low-quality requirements, and iteratively refines them to guide the generation of correct patches for resolving software issues. 
Figure~\ref{fig:overview} shows an overview of \tech{}, which consists of three main components: 
(1) \textbf{Requirement Generation} Component (Section~\ref{subsec:generation}) employs a requirement generation agent that autonomously explores the complex code repository to collect requirement-relevant context and systematically applies pre-defined requirement attributes to produce structured and information-rich issue-oriented requirements.
(2) \textbf{Requirement Assessment} Component (Section~\ref{subsec:assessment}) employs a requirement assessment agent to generate the initial patch (based on the constructed issue-oriented requirement), and leverages test execution to estimate patch correctness, thereby assessing the quality of this issue-oriented requirement.
(3) \textbf{Requirement Refinement} Component (Section~\ref{subsec:refinement}) employs a requirement analysis agent to diagnose the root causes if this issue-oriented requirement has quality deficiencies, and then provides actionable feedback to refine the requirement for more effective patch generation.
In the following sections, we introduce each component in detail. 
Here, we reuse the example introduced in Section~\ref{sec:example} to illustrate our approach.

\subsection{Requirement Generation}
\label{subsec:generation}

Prior research and established practices in software engineering have demonstrated that requirements engineering is a foundation of the software development lifecycle~\cite{rkaczkowska2023should, ramesh2021metrics, franch2023state}. 
This principle underscores the necessity of constructing high-quality requirements prior to implementation to ensure a thorough understanding of the software development task.
Inspired by principles from requirements engineering~\cite{pandey2010effective, jin2024mare, habiba2024mature}, we design a novel requirement generation agent that produces structured and information-rich issue-oriented requirements to guide patch generation (as shown in \texttt{Issue-oriented Requirement} of Figure~\ref{fig:example}). 
However, in issue resolution scenarios, automatically generating high-quality requirements remains challenging due to two key challenges.
First, accurately and effectively collecting issue-specific information is difficult, as such information is often scattered across multiple modules and files within large-scale codebases. 
Second, it is non-trivial to organize such fragmented information into a structured representation for effectively guiding patch generation.


To accurately and effectively collect issue-specific context, we design an effective requirements modeling strategy within the requirement generation agent. 
Following prior work~\cite{gao2025trae}, the requirement generation agent simulates the process of program comprehension by iteratively collecting and analyzing relevant code snippets in the codebase, thereby retrieving key contextual information. 
Starting from the issue description and associated code (as shown in \texttt{Original Issue Description} and \texttt{Codebase} of Figure~\ref{fig:example}), the agent progressively expands the context by retrieving additional code connected through program dependencies, using previously collected information to guide subsequent retrieval. 
To support this process, we equip the agent with an execution environment that provides access to the complete codebase and a suite of analysis tools (e.g., file retrieval, file browsing, and code analysis). 
Specifically, the agent autonomously invokes these tools within a customized Docker container, gathers the resulting outputs, and analyzes the feedback to determine subsequent actions. 
This iterative process continues until sufficient issue-specific context is obtained.


To address the challenge of requirement representation, we equip the requirement generation agent with a set of pre-defined requirement attributes (as shown in \texttt{Issue-oriented Requirement} of Figure~\ref{fig:example}),
enabling the transformation of previously collected, fragmented issue-related information into structured and comprehensive issue-oriented requirements.
Prior research in requirements engineering~\cite{kruchten20024+, inkermann2019model} indicates that complex software systems cannot be adequately described from a single perspective; instead, requirements should capture multiple dimensions, including system structure, functional behavior, and data interactions. 
Inspired by this multi-perspective abstraction principle~\cite{inkermann2019model, geisberger2007model, windisch2022approach}, we define nine primary requirement attributes along with seventeen corresponding sub-attributes, covering aspects ranging from overall codebase structure to fine-grained issue-specific details. 
The detailed requirement attributes are illustrated as follows:

\begin{mybreakablebox}
\begin{enumerate}[label=\arabic*., leftmargin=*, itemsep=2pt]
\item \textbf{Background:}
    \begin{itemize}[label=-, leftmargin=10pt, itemsep=1pt]
    \item Main Functionality: Describe the main purpose or capabilities of the repository.
    \item Main Modules: Describe the different modules corresponding to functionalities and the relationships between them.
    \end{itemize}

\item \textbf{Problem Overview:}
    \begin{itemize}[label=-, leftmargin=10pt, itemsep=1pt]
    \item Core Description: Provide a concise description of the problem.
    \item Problem Coverage: Indicate which functionalities and modules in the system are affected by the faulty code.
    \end{itemize}

\item \textbf{Steps to Reproduce:}
    \begin{itemize}[label=-, leftmargin=10pt, itemsep=1pt]
    \item Preconditions: Specify the required state, data, or configuration before reproducing the issue.
    \item Key Conditions: Clarify under what circumstances the error occurs, including specific inputs, versions, etc.
    \item Reproduction Commands: Provide the complete procedure from start to triggering the issue.
    \end{itemize}

\item \textbf{Actual Behavior:}
    \begin{itemize}[label=-, leftmargin=10pt, itemsep=1pt]
    \item Erroneous Behavior: Describe the actual erroneous output or exception from the faulty module.
    \item Correct Behavior: Describe the normal behavior of other modules using the affected code.
    \end{itemize}

\item \textbf{Expected Behavior:}
    \begin{itemize}[label=-, leftmargin=10pt, itemsep=1pt]
    \item Ideal Behavior: Describe the expected result when the system functions correctly.
    \item Success Criteria: Explain how to verify that the issue has been fixed.
    \end{itemize}

\item \textbf{Environment:}
    \begin{itemize}[label=-, leftmargin=10pt, itemsep=1pt]
    \item Dependencies and Imports: List the dependencies, required versions, APIs, libraries, or modules necessary to reproduce and fix the issue.
    \end{itemize}

\item \textbf{Root Cause Analysis:}
    \begin{itemize}[label=-, leftmargin=10pt, itemsep=1pt]
    \item Error Cause: Infer the potential root cause of the issue from the observed symptoms.
    \item Code Paths: Point out the key functions, call chains, or logic paths where the problem originates.
    \end{itemize}

\item \textbf{Solution:}
    \begin{itemize}[label=-, leftmargin=10pt, itemsep=1pt]
    \item Modification Location: Specify the files, functions, and code snippets that need to be modified.
    \item Modification Content: Provide a detailed description of the code snippets that need to be added, deleted, or modified, and explain specifically how the modifications should be made.
    \item Impact Scope: Explain which modules will be affected and which correct functionalities should remain unchanged.
    \end{itemize}

\item \textbf{Additional Notes:}
    \begin{itemize}[label=-, leftmargin=10pt, itemsep=1pt]
    \item Security, Compatibility, or Other Considerations: Provide additional notes that may affect future maintenance, deployment, or security.
    \end{itemize}
\end{enumerate}
\end{mybreakablebox}


Based on this standardized attribute schema, the requirement generation agent systematically organizes the collected contextual information and produces an information-rich and structured issue-oriented requirement, which serves as the foundation for subsequent patch generation.

\subsection{Requirement Assessment}
\label{subsec:assessment}


Software requirements engineering emphasizes that assessing the correctness of requirement specifications is a critical step~\cite{ramesh2021metrics, montgomery2022empirical, umar2024advances}, as low-quality requirements (e.g., incorrect or incomplete content) can lead to erroneous downstream implementations, such as inaccurate patch generation.
Traditional requirements engineering advocates the principle of Verification and Validation (V\&V)~\cite{freund2012ieee, terry2005requirements}, where verification examines whether requirements are correctly specified, consistent, and unambiguous, and validation assesses whether they can guide implementation toward the intended behavior.
However, existing requirement V\&V practices largely rely on manual effort~\cite{umar2024advances, franch2023state, bjarnason2014challenges}, and there is a lack of effective automatic methods for requirement assessment.
A fundamental challenge lies in the nature of requirements: they are typically expressed in structured natural language, which lacks formal semantics, making their correctness difficult to evaluate in a precise and quantitative manner~\cite{ferrari2014pragmatic, montgomery2022empirical, gigante2015semantic}.
To address this challenge, we design a requirement assessment agent that generates initial patches, thereby transforming requirement assessment into an evaluation of the resulting implementations. 
We then employ test execution to estimate patch correctness for inferring the quality of its underlying requirements.

Firstly, we explain the rationale for transforming the requirement quality assessment into the evaluation of the patch. 
Requirements engineering highlights the existence of traceability links between requirements and code, where requirements provide high-level abstractions of intended system behavior, and code represents their concrete implementations~\cite{mucha2024systematic, yoo2024building}. 
Despite differences in abstraction levels, both should be semantically consistent.
Motivated by this principle, we design a requirement assessment agent to generate executable patches based on requirements.

Next, we use the correctness of generated patches as a proxy for assessing the quality of requirements. 
The requirement assessment agent independently generates test scripts, each comprising: (1) \textit{reproduction tests}, which evaluate whether the generated patch resolves the target issue; and (2) \textit{regression tests}, which verify that the patch does not introduce unintended side effects on existing functionality. 
Specifically, reproduction tests are constructed based on the issue-oriented requirements (\textit{Reproduction Commands} and \textit{Success Criteria} requirement attributes). 
Although original codebases typically include existing regression tests, prior work~\cite{xia2025demystifying, tang2015will, tian2026agent} has shown that issue resolution may legitimately modify certain behaviors, potentially causing some existing regression tests to fail. 
To address this, the agent generates refined regression tests by leveraging both the issue-oriented requirements (\textit{Correct Behavior} requirement attributes) and the existing (potentially outdated or inconsistent) regression tests in the original codebase. 
Furthermore, to maximize test diversity while controlling computational cost, we adopt a high-temperature sampling strategy~\cite{shur2024growing, zhang2021trading, minhturning} to generate ten test scripts for each issue.
We acknowledge that, due to the inherent limitations of LLM agents (e.g., hallucination), the generated tests may contain inaccuracies, and thus perfect test quality cannot be guaranteed. 
Nevertheless, consistent with prior work~\cite{xia2025demystifying, ruan2025specrover}, such tests still provide a useful overall evaluation signal, enabling the selection of correct patches based on majority voting criteria.
We further validate this observation in our ablation study (Section~\ref{subsec:rq3}).
To further mitigate the impact of potentially faulty tests, we iteratively repair and refine test scripts based on updated issue-oriented requirements in each iteration (Section~\ref{subsec:refinement}). 
A detailed discussion of test quality will be presented in Section~\ref{subsec:testcase}.


To quantify the requirement quality, we introduce the \textbf{Requirement Assessment Score} (RAS), defined as the ratio of test scripts passed by the generated patch to the total number of test scripts.
The RAS ranges from 0 to 1.0, where higher values indicate that the issue-oriented requirement guides correct implementation more effectively, thereby reflecting higher requirement quality. 
We adopt a strict acceptance criterion, i.e., a patch is considered fully compliant with the requirement only when its RAS equals 1.0.
In such cases, the patch is directly accepted as the final output. 
Conversely, an RAS below 1.0 suggests potential deficiencies in the requirement (e.g., incomplete information or inaccurate descriptions), which may lead to errors in the generated patches or tests.
Accordingly, requirements associated with low RAS values are passed to the subsequent requirement refinement component for further improvement.

\subsection{Requirement Refinement}
\label{subsec:refinement}

This component aims to guide LLMs in generating correct patches by iteratively refining low-quality requirements. 
However, accurately localizing requirement deficiencies in low-quality requirements and performing targeted refinements are challenging due to two key limitations. 
First, the large semantic space of requirements (characterized by complex requirement attributes and extensive requirement expressions) makes it difficult to precisely pinpoint root causes. 
Second, effective refinement requires actionable guidance to support targeted optimization; however, such feedback mechanisms remain underdeveloped, limiting the ability to systematically improve low-quality requirements.


To address the challenge of investigating requirement deficiencies, we draw on the \textit{IEEE 830 standard}~\cite{doe2011recommended} from requirements engineering and simplify the diverse range of requirement deficiencies in issue resolution tasks into three categories: \textit{Conflict}, \textit{Omission}, and \textit{Ambiguity}.
These categories are designed to be mutually complementary and collectively comprehensive, which enables focusing on a limited set of deficiency patterns for low-quality requirements, thereby reducing the search space of root cause identification.
Specifically, we employ a requirement analysis agent and design targeted system prompt cues to guide the agent in determining potential deficiency categories in low-quality requirements.
The agent is allowed to assign one or multiple deficiency categories for each low-quality requirement. 
Detailed descriptions of these deficiency categories are as follows:

\begin{mybreakablebox}
    \begin{enumerate}[label=\arabic*., leftmargin=*, itemsep=2pt]
    \item \textbf{Conflict}: This deficiency category concerns the correctness of issue-oriented requirements, characterized by inconsistencies between the requirements and the issue description, such that the requirements fail to accurately reflect the true intent of the problem.
    
    \item \textbf{Omission}: This deficiency category concerns the completeness of issue-oriented requirements, characterized by missing key information described in the issue, resulting in requirements that fail to comprehensively specify the intended behaviors or constraints.
    
    \item \textbf{Ambiguity}: This deficiency category concerns the clarity of issue-oriented requirements, characterized by vague or unclear descriptions that may lead to different interpretations by different agents, thereby affecting requirement executability and consistency.
    \end{enumerate}
\end{mybreakablebox}

To address the challenge of designing effective feedback mechanisms, we design tailored refinement strategies for each category of requirement deficiencies and employ the requirement analysis agent to generate actionable feedback accordingly. 
Note that issue descriptions, codebase, requirements, patches, and test scripts collectively constitute the complete set of task inputs and contextual information in \tech{}. 
Consequently, requirement deficiencies may manifest in different forms across these elements.
Based on this unified view, we design three category-specific refinement strategies:
(1) For \textit{conflict} deficiency, the agent first evaluates their semantic correctness against the original issue description. 
It then cross-validates the requirements using the generated patches, test scripts, and relevant code context retrieved from the codebase. 
Based on this analysis, the agent produces detailed error diagnostics (e.g., incorrect requirement attributes or representations) as corresponding refinement guidelines to support requirement correction.
(2) For \textit{omission} deficiency, the agent identifies missing information by comparing the requirements with the original issue description. 
It then determines the requirement attributes associated with the missing information and their impact on behavioral correctness, and formulates refinement guidelines to explicitly capture the omitted details.
(3) For \textit{ambiguity} deficiency, the agent examines semantic inconsistencies between the requirements and their corresponding patches and test scripts to detect ambiguous expressions.
It then determines their intended meaning and generates refinement guidelines to improve requirement clarity. 
All analyses are automatically performed by the requirement analysis agent under pre-defined system prompts and tool support. 
The agent then aggregates the identified requirement deficiency categories and their corresponding refinement guidelines into structured feedback, which is used to support iterative requirement refinement.

Specifically, during the iterative process,
we adopt a greedy optimization strategy~\cite{garcia2025greedy}, which selects the best candidate requirement in each iteration based on Requirement Assessment Score (introduced in Section~\ref{subsec:assessment}).
This prioritizes the higher-quality requirement and incorporates it with the generated feedback as input to the requirement generation agent (introduced in Section~\ref{subsec:generation}) for subsequent iterations. 
In each iteration, the requirement generation agent updates the issue-oriented requirement based on the provided feedback, enabling progressive refinement. 
To further enhance feedback effectiveness, the system records non-improving feedback as counterexamples, encouraging the requirement analysis agent to adjust its refinement guidelines in subsequent iterations.
The number of iterations (i.e., \textit{$N$}), which serves as a key hyper-parameter for balancing effectiveness and computational overhead, is analyzed in detail in Section~\ref{subsec:rq2}. When $N$ reaches the maximum number of iterations, \tech{} generates the final patch as the output based on the requirement with the highest RAS, ensuring that the final patch achieves the best quality within the current iterative process.
Through this component, \tech{} effectively refines issue-oriented requirements to facilitate the generation of correct patches (as shown in \texttt{Correct Patch} of Figure~\ref{fig:example}).

\section{Evaluation Design}
\label{sec:design}

This work focuses on the following research questions (RQs):
\begin{itemize}[leftmargin=10pt]
    \item \textbf{RQ1:} How does \tech{} perform in terms of effectiveness and efficiency compared to the state-of-the-art techniques?
    \item \textbf{RQ2:} How do hyper-parameters affect \tech{}'s effectiveness?
    \item \textbf{RQ3:} How does each main component in \tech{} contribute to the overall effectiveness?
\end{itemize}

\subsection{Benchmarks}
\label{subsec:benchmark}

Following prior work~\cite{yang2024swe, xia2025demystifying, zhang2024autocoderover, wangopenhands}, we evaluate the performance of \tech{} on SWE-bench~\cite{jimenez2024swe}, a widely-used benchmark for software issue resolution.
Specifically, we consider two \textit{subsets} and one \textit{extension} of SWE-bench: \textbf{SWE-bench Lite}~\cite{jimenez2024swe}, \textbf{SWE-bench Verified}~\cite{openai_swebench_verified}, and \textbf{SWE-bench Pro}~\cite{deng2025swe}.

\begin{itemize}[leftmargin=10pt]
    \item \textbf{SWE-bench Lite}~\cite{jimenez2024swe} is a lightweight subset of SWE-bench, designed for efficient evaluation. 
    It contains 300 GitHub issues and covers 11 out of the 12 repositories in the original SWE-bench dataset, while preserving a similar cross-repository distribution and diversity of issues. 

    \item \textbf{SWE-bench Verified}~\cite{openai_swebench_verified} is a curated, high-quality subset of SWE-bench. 
    To reduce noisy issues, OpenAI~\cite{openai} curated 500 high-quality instances through execution-based validation and manual verification, providing a more reliable benchmark.

    \item \textbf{SWE-bench Pro}~\cite{deng2025swe} is an extension of SWE-bench constructed by Scale AI~\cite{scaleai}, comprising 731 instances across multiple repositories and programming languages. 
    It reflects more realistic industrial scenarios and introduces greater complexity, thereby enabling the evaluation of generalization in multi-language and large-scale codebases.
\end{itemize}

Due to the high computational cost and long runtime of repository-level issue resolution tasks, we follow the prior work~\cite{lindenbauer2025complexity} and sample 100 instances from each benchmark for evaluation. 
For SWE-bench Lite and SWE-bench Verified, we ensure that the sampled instances cover all repositories included in the respective datasets. 
As these two benchmarks primarily focus on Python, we exclude Python-related issues when sampling from SWE-bench Pro, allowing us to better evaluate the performance of \tech{} in multi-language repository environments.

\subsection{Metrics}
\label{subsec:metric}

Following prior work~\cite{yang2024swe,tao2024magis,jimenez2024swe,zanmulti}, we evaluate the effectiveness of \tech{} using \textit{\% Applied} and \textit{\% Resolved}, and measure its efficiency using \textit{\# Input Tokens}, \textit{\# Output Tokens}, and \textit{\$ Cost}.

\begin{itemize}[leftmargin=10pt]    
    \item \textbf{\textit{\% Applied}} measures the syntactic correctness of generated patches by determining whether a patch can be successfully applied to the codebase. 
    Specifically, a patch is considered syntactically correct if it can be applied using \texttt{git apply} without errors. 
    \textit{\% Applied} is computed as the percentage of such successfully applied patches over all generated patches, where a higher value indicates better syntactic validity.
    
    \item \textbf{\textit{\% Resolved}} measures the functional correctness of generated patches by assessing whether they successfully resolve the target software issues. 
    A patch is considered correct if it passes all associated golden test cases provided by the benchmark. 
    \textit{\% Resolved} is defined as the percentage of issues successfully resolved out of the total number of issues, with higher values indicating better issue resolution performance.
    
    \item \textbf{\textit{\# Input Tokens}} and \textbf{\textit{\# Output Tokens}} denote the average number of tokens in the LLM input prompts and generated outputs, respectively, for solving a single software issue. 
    These metrics capture the token-level overhead of a technique in practical deployment, where lower values indicate higher efficiency.
    
    \item \textbf{\textit{\$ Cost}} represents the average monetary cost of LLM API inference required to resolve a single software issue. 
    Lower cost values indicate higher efficiency.
\end{itemize}

\subsection{Baselines}
\label{subsec:baseline}



To comprehensively evaluate \tech{}, we consider three state-of-the-art or representative automated issue resolution techniques:

\begin{itemize}[leftmargin=10pt]    
    \item \textbf{BM25 Retrieval}~\cite{jimenez2024swe} is a classical retrieval-augmented approach introduced in SWE-bench for issue resolution. 
    Specifically, it employs the BM25 (Best Matching 25)~\cite{robertson1994okapi} algorithm to retrieve issue-relevant context from the codebase for bug localization, and subsequently generates patches directly based on the retrieved content, forming a standard retrieve-generate pipeline.

    \item \textbf{Agentless}~\cite{xia2025demystifying} is a state-of-the-art workflow-based technique for automated software issue resolution. 
    Specifically, it follows a manually predefined three-phase process (i.e., localization, patch generation, and patch validation) without relying on autonomous agents to plan actions or interact with complex tools.
    
    \item \textbf{Trae-agent}~\cite{gao2025trae} is a state-of-the-art industrial agent designed for general-purpose software engineering tasks. 
    It provides a powerful command-line interface (CLI) capable of accurately interpreting natural language instructions and automatically executing complex workflows through various agent tools. 
    To ensure a fair comparison and control computational cost, we disable its test-time scaling module.
\end{itemize}


To the best of our knowledge, \tech{} is the first approach to introduce a requirements-driven strategy for software issue resolution. 
To further investigate this dimension, we additionally adapt two state-of-the-art baselines from function-level code generation that focus on requirement completion and alignment:

\begin{itemize}[leftmargin=10pt]
    \item \textbf{ArchCode}~\cite{han2024archcode} is a state-of-the-art approach for requirement completion in function-level code generation. 
    It identifies incomplete requirements by inferring missing behavioral and performance constraints, and augments them from both functional and non-functional perspectives to improve code generation quality.
    
    \item \textbf{Specine}~\cite{tian2025aligning} represents the state of the art in requirement alignment for function-level code generation. 
    It leverages a domain-specific language (DSL) to explicitly model discrepancies between requirement specifications and generated code, and iteratively refines requirement descriptions to better align the model's understanding with the intended semantics, thereby improving code correctness.
\end{itemize}

It is important to note that both ArchCode and Specine do not encompass a complete requirements engineering lifecycle; rather, they focus solely on completing or aligning existing requirements, with optimization strategies tailored to function-level code generation. 
These characteristics distinguish them from our approach. 
Moreover, as ArchCode and Specine are originally designed for function-level tasks, they lack mechanisms for retrieving contextual information from the codebase.
To adapt these methods to repository-level tasks, we integrate the BM25 Retrieval approach into both approaches to obtain relevant context, thereby enabling patch generation in a repository setting. 


\subsection{Implementation Details} 
\label{subsec:detail}

To evaluate the performance of \tech{}, we select two advanced LLMs as base models: \texttt{DeepSeek-V3.2}~\cite{liu2025deepseek} and \texttt{Qwen-Plus}~\cite{yang2025qwen3}. 
Both models have demonstrated strong capabilities in software engineering tasks (e.g., code generation and testing) and have been widely adopted in prior studies~\cite{wang2025exploring, mahran2025investigating, chen2026beyondswe}. 
These models are accessed via APIs provided by \textit{DeepSeek} and \textit{Alibaba Cloud}, respectively.
Regarding experimental settings, to reduce randomness during generation while preserving a certain degree of diversity, we set the LLM temperature to 0.1 for patch generation. 
In contrast, to encourage greater diversity during requirement generation, refinement and test generation, we set the temperature to 0.5. 
Considering the trade-off between cost and effectiveness, and following prior work~\cite{han2026tdflow}, we set the maximum number of iterations (i.e., \textit{N}) for all studied iterative techniques to 4 to ensure a fair comparison. 
Furthermore, we investigate the impact of $N$ in Section~\ref{subsec:rq2}.
In addition, considering the trade-off between computational cost and effectiveness, same as prior work~\cite{deng2025swe}, we limit the maximum number of agent interaction turns to 50 for all techniques.

\section{Results and Analysis}
\label{sec:results}
\subsection{RQ1: Effectiveness and Efficiency}
\label{subsec:rq1}

\begin{table}[t!]
    \caption{Effectiveness comparison in terms of \textit{\% Applied} ($\uparrow$) and \textit{\% Resolved} ($\uparrow$).}
    \vspace{-2mm}
    \label{tab:RQ1-effectiveness} 
    \centering
    \setlength{\tabcolsep}{1.6mm}
    \normalsize
    \resizebox{\linewidth}{!}{%
    \begin{tabular}{lcccccc}
    \toprule
    \multirow{2}{*}{\textbf{Tech.}}
    & \multicolumn{2}{c}{\textbf{SWE-Lite}}
    & \multicolumn{2}{c}{\textbf{SWE-Verified}}
    & \multicolumn{2}{c}{\textbf{SWE-Pro}} \\
    \cmidrule(lr){2-3} \cmidrule(lr){4-5} \cmidrule(lr){6-7}
    & \% App. & \% Res.
    & \% App. & \% Res.
    & \% App. & \% Res. \\
    \midrule
    \multicolumn{7}{l}{\cellcolor{gray!30}\textbf{DeepSeek-V3.2}} \\
    \midrule
    BM25       & 34\% & 6\%  & 35\% & 7\%  & 14\% & 3\% \\
    Agentless  & 55\% & 24\% & 61\% & 35\% & 62\% & 6\% \\
    Trae-agent & 64\% & 28\% & 61\% & 35\% & 43\% & 11\% \\
    Specine    & 52\% & 13\% & 47\% & 11\% & 27\% & 4\% \\
    ArchCode   & 32\% & 14\% & 22\% & 11\% & 31\% & 6\% \\
    \textbf{\tech{}} & \textbf{75\%} & \textbf{37\%} & \textbf{83\%} & \textbf{46\%} & \textbf{75\%} & \textbf{21\%} \\
    \midrule
    \multicolumn{7}{l}{\cellcolor{gray!30}\textbf{Qwen-Plus}} \\
    \midrule
    BM25       & 30\% & 4\%  & 32\% & 4\%  & 26\% & 2\% \\
    Agentless  & 47\% & 14\% & 35\% & 22\% & 59\% & 5\% \\
    Trae-agent & 55\% & 17\% & 63\% & 24\% & 49\% & 5\% \\
    Specine    & 40\% & 7\%  & 34\% & 10\%  & 57\% & 4\% \\
    ArchCode   & 33\% & 6\%  & 29\% & 7\%  & 49\% & 4\% \\
    \textbf{\tech{}} & \textbf{85\%} & \textbf{24\%} & \textbf{80\%} & \textbf{32\%} & \textbf{70\%} & \textbf{15\%} \\
    \bottomrule
    \multicolumn{7}{l}{\footnotesize * Bold values indicate the best performance among all techniques;} \\
    \multicolumn{7}{l}{\footnotesize * \% App. and \% Res. are the abbreviations of \% Applied and \% Resolved, respectively;} \\ 
    \multicolumn{7}{l}{\footnotesize * SWE-Lite, SWE-Verified, SWE-Pro are the abbreviations of SWE-bench Lite, SWE-} \\ 
    \multicolumn{7}{l}{\footnotesize  \,\, bench Verified, SWE-bench Pro, respectively.}
    \end{tabular}%
    }
\end{table}

\subsubsection{Process:} 

To address RQ1, we apply \tech{} and five baselines (i.e., BM25 Retrieval, Agentless, Trae-agent, ArchCode, and Specine) on two advanced LLMs (i.e., DeepSeek-V3.2 and Qwen-Plus). 
We evaluate their effectiveness on three widely-used benchmarks (i.e., SWE-bench Lite, SWE-bench Verified, and SWE-bench Pro) using the \textit{\% Resolved} and \textit{\% Applied} metrics. 
In addition, we report the number of uniquely resolved issues achieved by each technique to further assess their complementary strengths.
For efficiency evaluation, we measure the computational overhead of each technique in terms of \textit{\# Input Tokens}, \textit{\# Output Tokens}, and \textit{\$ Cost}. 
Notably, these efficiency metrics are computed over resolved issues only, enabling a more precise analysis of the cost required to successfully resolve an issue.

\subsubsection{Results:}


\begin{table}[t!]
    \caption{Efficiency comparison in terms of \textit{\# Input Tokens} ($\downarrow$), \textit{\# Output Tokens} ($\downarrow$), and \textit{\$ Cost} ($\downarrow$).}
    \vspace{-2mm}
    \label{tab:RQ1-efficiency} 
    \centering
    \tabcolsep=0.7mm
    \normalsize
    \begin{tabular}{lcccccc}
    \toprule
    \multirow{2}{*}{\textbf{Tech.}}
    & \multicolumn{2}{c}{\textbf{\# Input Tokens}}
    & \multicolumn{2}{c}{\textbf{\# Output Tokens}}
    & \multicolumn{2}{c}{\textbf{\$ Cost}} \\
    \cmidrule(lr){2-3} \cmidrule(lr){4-5} \cmidrule(lr){6-7}
    & DeepSeek & Qwen
    & DeepSeek & Qwen
    & DeepSeek & Qwen \\
    \midrule
    BM25             & 0.34M & 0.51M & 0.02M & 0.03M & 0.102 & 0.068 \\
    Agentless        & 1.33M & 0.97M & 0.12M & 0.09M & 0.424 & 0.138 \\
    Trae-agent       & 3.67M & 3.19M & 0.04M & 0.03M & 1.046 & 0.379 \\
    Specine          & 1.53M & 1.96M & 0.10M & 0.17M & 0.470 & 0.276 \\
    ArchCode         & 2.34M & 6.56M & 0.24M & 0.60M & 0.757 & 0.934 \\
    \textbf{\tech{}} & 5.13M & 3.21M & 0.09M & 0.05M & 1.474 & 0.386 \\
    \bottomrule
    \end{tabular}
\end{table}



\begin{figure*}[t!]
    \centering
    \includegraphics[width=\textwidth]{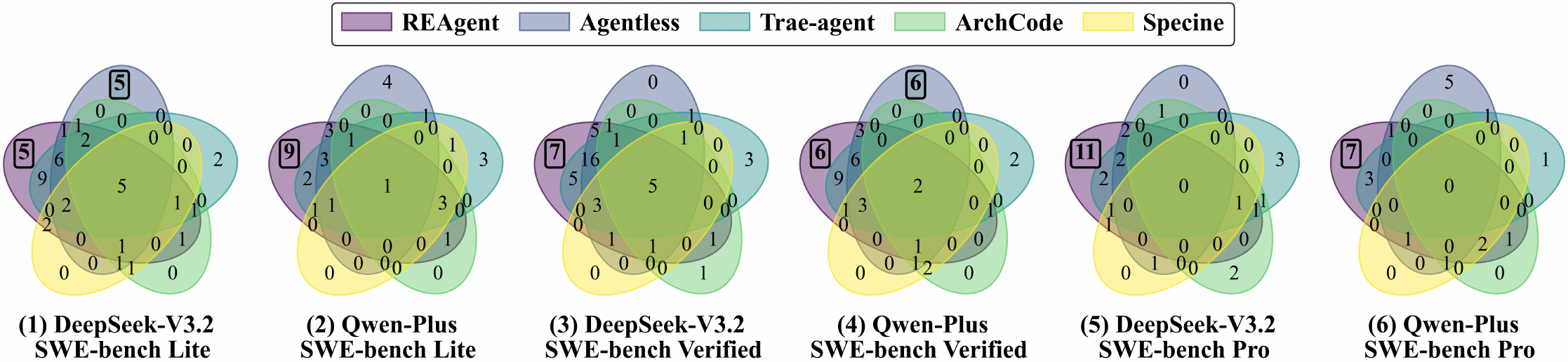}
    \caption{Number of uniquely resolved instances across different techniques}
    \label{fig:venn}
\end{figure*}



Table~\ref{tab:RQ1-effectiveness} presents the effectiveness comparison across all techniques. 
We observe that \tech{} consistently achieves the best performance among all baselines, demonstrating superior results across both LLMs and all three benchmarks. 
In particular, \tech{} improves over existing methods by 9.17\%$\sim$24.83\% and 22.17\%$\sim$49.50\% in terms of \textit{\% Resolved} and \textit{\% Applied}, respectively. 
Furthermore, the \textit{Wilcoxon Signed-Rank Test}~\cite{1963Critical} (at a significance level of 0.05) yields p-values below $2.5 \times 10^{-4}$, demonstrating the statistically significant superiority of \tech{} over all baselines in terms of \textit{\% Resolved} and \textit{\% Applied}. 
Figure~\ref{fig:venn} further illustrates the overlap of successfully resolved issues via a Venn diagram. 
We observe that \tech{} yields the largest number of uniquely resolved issues, indicating that it not only outperforms existing approaches overall but also complements them by solving cases that others fail to address. 
Collectively, these results demonstrate the significant effectiveness of \tech{} in enhancing LLM-based issue resolution. 
Additionally, to assess the risk of potential data leakage, we analyze the number of patches generated by \tech{} that are textually identical to the corresponding gold patches. 
The results show that \tech{} produces only two identical patches across all experiments, mitigating the threats of data leakage.
Besides, the authors manually inspected all correct patches to further verify their correctness and to ensure that none of the results were obtained through reward hacking.

Table~\ref{tab:RQ1-efficiency} reports the efficiency of each technique, measured as the average cost of successfully resolving an issue. 
We find that \tech{} generally incurs higher \textit{\# Input Tokens}, \textit{\# Output Tokens}, and \textit{\$ Cost} compared to BM25 Retrieval, Agentless, Trae-agent, and Specine, while remaining comparable to ArchCode.
Specifically, \tech{} incurs a token overhead of approximately 5.13M (DeepSeek-V3.2) and 3.21M (Qwen-Plus) for \textit{\# Input Tokens}, and 0.09M and 0.05M for \textit{\# Output Tokens}, corresponding to \textit{\$ Costs} of 1.474 and 0.386, respectively. 
Although \tech{} is not the most efficient method, its substantial effectiveness gains justify the additional computational overhead, indicating a favorable trade-off between effectiveness and efficiency. 
In practice, the associated cost is indeed acceptable for real-world deployment. 
Furthermore, we provide the number of iterations ($N$), which can be adjusted by developers to balance effectiveness and efficiency in practice. 
As further analyzed in Section~\ref{subsec:rq2}, even when increasing $N$ for baseline techniques to incur higher costs than \tech{}, \tech{} still achieves the best performance, highlighting its superior effectiveness and scalability.
We discuss potential directions for improving efficiency (e.g., more efficient context management) in Section~\ref{subsec:future_work} as part of future work.


\begin{figure*}[t!]
    \centering
    \includegraphics[width=\textwidth]{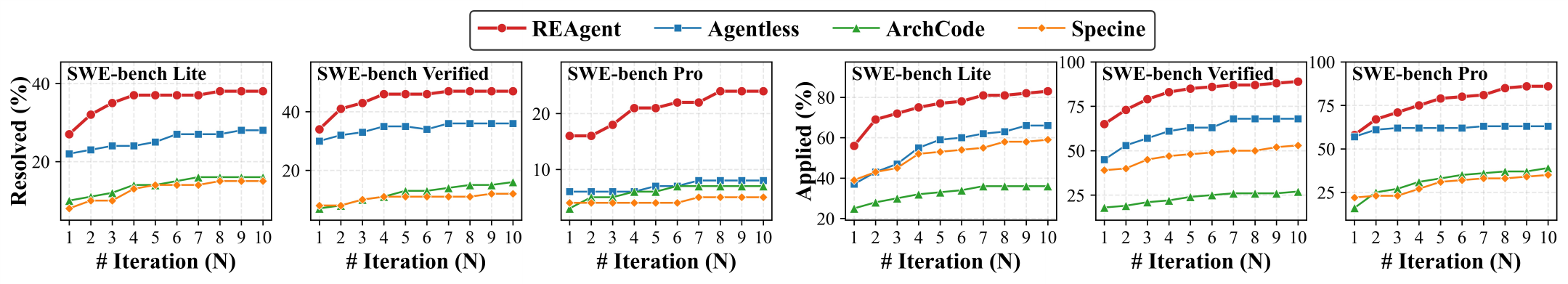}
    \caption{Influence of the number of iterations ($N$) on \textit{\% Applied} ($\uparrow$) and \textit{\% Resolved} ($\uparrow$) using DeepSeek-V3.2}
    \label{fig:rq2}
\end{figure*}

\subsection{RQ2: Influence of Hyper-parameter}
\label{subsec:rq2}
\subsubsection{Process:}

The number of iterations ($N$) is a critical hyper-parameter for all iterative techniques. 
In this RQ, we investigate its impact on the effectiveness of \tech{} and other iterative baselines (i.e., Agentless, ArchCode, and Specine).
To control experimental cost while ensuring meaningful comparisons, we adopt the more effective LLM DeepSeek-V3.2 as the base model for all evaluated techniques in this RQ. Specifically, we vary the number of iterations within the range $1 \le N \le 10$, and compare the effectiveness of \tech{} and the baselines using the \textit{\% Applied} and \textit{\% Resolved} metrics.

\subsubsection{Results:}

\begin{table}[t!]
    \caption{Comparison of effectiveness and efficiency between \tech{} ($N=1$) and baselines ($N=10$) using DeepSeek-V3.2.}
    \vspace{-2mm}
    \label{tab:comparable} 
    \centering
    \tabcolsep=1.8mm
    \normalsize
    \begin{tabular}{lccccc}
    \toprule
    \textbf{Technique} &
    \textbf{\makecell[c]{\# Input \\ Tokens}} & 
    \textbf{\makecell[c]{\# Output \\ Tokens}} & 
    \textbf{\$ Cost} &
    \textbf{\% Resolved}  \\
    \midrule
    Agentless$_{N=10}$                 & \textbf{1.72M} & 0.23M & 0.577 & 24.00\% \\
    Specine$_{N=10}$                   & 3.39M & 0.19M & 1.029 & 10.67\% \\
    ArchCode$_{N=10}$                  & 4.65M & 0.48M & 1.503 & 13.00\% \\ \midrule
    \textbf{\tech{}$_{N=1}$}  & 1.73M & \textbf{0.06M} & \textbf{0.510} & \textbf{25.67\%} \\
    \bottomrule
    \end{tabular}
\end{table}



Figure~\ref{fig:rq2} illustrates the performance trends of different techniques as the number of iterations ($N$) increases.
First, we observe that increasing $N$ consistently improves the performance of all techniques on both metrics. 
Notably, \tech{} consistently outperforms all baselines across all settings. 
On average, \tech{} achieves improvements of 6.33\%$\sim$26.00\% in terms of \textit{\% Resolved} and 13.33\%$\sim$52.33\% in terms of \textit{\% Applied} compared to baseline techniques, demonstrating stable and substantial advantages under varying iteration budgets.


Furthermore, \tech{} benefits more significantly from the increased number of iterations than the baselines. 
As $N$ increases from 1 to 10, \tech{} achieves an average improvement of 10.67\% in terms of \textit{\% Resolved} across the three benchmarks, whereas the baselines exhibit smaller gains of 4.00\%$\sim$6.33\%. 
A similar trend is observed for \textit{\% Applied}, where \tech{} improves by 26.33\% on average, compared to 14.33\%$\sim$19.33\% for baselines. 
These results indicate that \tech{} demonstrates stronger capability in leveraging additional iterations for continuous performance improvement. 


Table~\ref{tab:comparable} shows an extreme comparison setting in which \tech{} is configured with $N=1$, while all baselines are configured with $N=10$, allowing for a joint evaluation of effectiveness and efficiency. 
Even under this constrained setting, \tech{} remains superior to all baselines, achieving an average \textit{\% Resolved} of 25.67\%. 
Moreover, the average cost per successfully resolved issue for \tech{} is significantly lower than that of the baselines. 
These findings suggest that \tech{} achieves a favorable balance between effectiveness and efficiency, and that its performance can be further enhanced with additional computational budget.

\subsection{RQ3: Contribution of Main Components}
\label{subsec:rq3}
\subsubsection{Variants:}

\tech{} consists of three core components: requirement generation, requirement assessment, and requirement refinement. 
To systematically evaluate the contribution of each component, we construct four ablated variants of \tech{}.

For the requirement generation component, \tech{} adopts a requirement modeling strategy to collect issue-relevant context and defines structured requirement attributes to represent issue-oriented requirements in a standardized manner. 
Based on this design, we construct two variants:
\textbf{\tech{}$_{woRM}$}, which replaces the requirement modeling strategy with a BM25-based retrieval strategy for context collection;
and \textbf{\tech{}$_{woRA}$}, which removes the predefined requirement attributes and instead prompts the LLM to generate requirements in an unstructured manner.

For the requirement assessment component, \tech{} employs a requirement assessment agent to generate an initial patch based on the constructed requirements, and leverages test execution to estimate patch correctness, thereby assessing requirement quality.
To evaluate its contribution, we construct a variant denoted as \textbf{\tech{}$_{woA}$}, which replaces this component with a widely-used \textit{LLM-as-a-judge} strategy to directly assess requirement quality.

For the requirement refinement component, \tech{} utilizes a requirement analysis agent to diagnose root causes of requirement deficiencies and provide effective feedback for subsequent requirement refinement and patch generation. 
To investigate its impact, we construct a variant \textbf{\tech{}$_{woR}$}, which replaces this component with a commonly used test execution-based feedback strategy to iteratively repair requirements and patches.

\begin{table}[t!]
    \caption{The effectiveness comparison of \tech{} and its five variants in terms of \textit{\% Applied} ($\uparrow$) and \textit{\% Resolved} ($\uparrow$).}
    \vspace{-2mm}
    \label{tab:RQ3} 
    \centering
    \setlength{\tabcolsep}{1.6mm}
    \normalsize
    \resizebox{\linewidth}{!}{%
    \begin{tabular}{lcccccc}
    \toprule
    \multirow{2}{*}{\textbf{Variant}}
    & \multicolumn{2}{c}{\textbf{SWE-Lite}}
    & \multicolumn{2}{c}{\textbf{SWE-Verified}}
    & \multicolumn{2}{c}{\textbf{SWE-Pro}} \\
    \cmidrule(lr){2-3} \cmidrule(lr){4-5} \cmidrule(lr){6-7}
    & \% App. & \% Res.
    & \% App. & \% Res.
    & \% App. & \% Res. \\
    \midrule
    \multicolumn{7}{l}{\cellcolor{gray!30}\textbf{DeepSeek-V3.2}} \\
    \midrule
    \tech{}$_{woRM}$     & 69\% & 28\% & 74\% & 31\% & 63\% & 12\% \\
    \tech{}$_{woRA}$       & 70\% & 34\% & 78\% & 40\% & 72\% & 18\% \\
    \tech{}$_{woA}$      & 49\% & 26\% & 58\% & 34\% & 40\% & 14\% \\
    \tech{}$_{woR}$      & 72\% & 35\% & 79\% & 44\% & 74\% & 19\% \\
    \textbf{\tech{}}         & \textbf{75\%} & \textbf{37\%} & \textbf{83\%} & \textbf{46\%} & \textbf{75\%} & \textbf{21\%} \\
    \midrule
    \multicolumn{7}{l}{\cellcolor{gray!30}\textbf{Qwen-Plus}} \\
    \midrule
    \tech{}$_{woRM}$     & 79\% & 13\% & 78\% & 25\% & 63\% & 9\% \\
    \tech{}$_{woRA}$       & 81\% & 21\% & 72\% & 29\% & 58\% & 13\% \\
    \tech{}$_{woA}$      & 68\% & 18\% & 67\% & 26\% & 38\% & 11\% \\
    \tech{}$_{woR}$      & 82\% & 21\% & 77\% & 30\% & 64\% & 12\% \\
    \textbf{\tech{}}         & \textbf{85\%} & \textbf{24\%} & \textbf{80\%} & \textbf{32\%} & \textbf{70\%} & \textbf{15\%} \\
    \bottomrule
    \end{tabular}%
    }
\end{table}

\subsubsection{Results:}

Table~\ref{tab:RQ3} presents the comparison between \tech{} and its four variants across both LLMs and three benchmarks, evaluated using \textit{\% Resolved} and \textit{\% Applied} metrics.
First, \tech{} consistently outperforms \tech{}$_{woRM}$ and \tech{}$_{woRA}$ on both metrics. 
On average, \tech{} achieves improvements of 9.50\% and 3.33\% over \tech{}$_{woRM}$ and \tech{}$_{woRA}$ in terms of \textit{\% Resolved}, respectively, as well as gains of 7.00\% and 6.17\% in terms of \textit{\% Applied}. 
These results validate the effectiveness of both the requirement modeling strategy and the structured requirement attributes adopted in the requirement generation component of \tech{}.


Second, \tech{} significantly outperforms \tech{}$_{woA}$, with average improvements of 7.67\% in \textit{\% Resolved} and 24.67\% in \textit{\% Applied}. 
Notably, \tech{}$_{woA}$ exhibits the weakest performance among all variants, underscoring the importance of accurate requirement assessment. 
This finding indicates that reliable requirement assessment substantially improves downstream requirement refinement and patch generation. 
Moreover, since this component relies on generated test scripts, the results also suggest that the generated tests are of reasonably high quality, highlighting the practicality of \tech{} in real-world scenarios where test cases may be incomplete or unavailable. 
We further analyze the quality of generated tests in Section~\ref{subsec:testcase} and discuss potential improvements in Section~\ref{subsec:future_work}.


Third, \tech{} outperforms \tech{}$_{woR}$, achieving average improvements of 2.33\% in \textit{\% Resolved} and 3.33\% in \textit{\% Applied}. 
These results demonstrate the contribution of the requirement refinement component and further indicate that it is more effective than the widely used test execution-based feedback strategy. 

Furthermore, the \textit{Wilcoxon Signed-Rank Test}~\cite{1963Critical} (at a significance level of 0.05) yields p-values below $2.5 \times 10^{-4}$, demonstrating the statistically significant superiority of \tech{} over all variants in terms of \textit{\% Resolved} and \textit{\% Applied}.

\section{Discussion}
\label{sec:discussion}

\begin{table}[t!]
    \caption{The correctness of LLM-generated tests.}
    \vspace{-2mm}
    \label{tab:testcase} 
    \centering
    \tabcolsep=0.85mm
    \small
    \begin{tabular}{lcccccc}
    \toprule
    \multirow{2}{*}{\textbf{ Tech. }}
    & \multicolumn{2}{c}{\textbf{SWE-Lite}}
    & \multicolumn{2}{c}{\textbf{SWE-Verified}}
    & \multicolumn{2}{c}{\textbf{SWE-Pro}} \\
    \cmidrule(lr){2-3} \cmidrule(lr){4-5} \cmidrule(lr){6-7}
    & DeepSeek & Qwen
    & DeepSeek & Qwen
    & DeepSeek & Qwen \\
    \midrule
    \textit{Base}   & 15.98\% & 15.12\% & 12.92\% & 13.25\% & 10.55\% & 10.37\%  \\
    \tech{}$_{N=1}$ & 26.20\% & 22.10\% & 19.91\% & 27.38\% & 14.97\% & 23.91\% \\
    \tech{}$_{N=2}$ & 36.20\% & 34.12\% & 28.02\% & 36.10\% & 21.10\% & 35.70\% \\
    \tech{}$_{N=3}$ & 42.20\% & 41.68\% & 32.11\% & 43.39\% & 27.01\% & 40.50\%  \\
    \tech{}$_{N=4}$ & \textbf{46.00\%} & \textbf{56.14\%} & \textbf{36.84\%} & \textbf{47.70\%} & \textbf{30.50\%} & \textbf{44.04\%} \\
    \bottomrule
    \end{tabular}
\end{table}

\subsection{Quality of LLM-Generated Tests}
\label{subsec:testcase}

The quality of LLM-generated tests plays a critical role in \tech{}. 
To comprehensively evaluate test quality, we assess their correctness with respect to the gold patches provided in the benchmarks. 
Specifically, a test script is considered correct if reproduction tests can successfully reproduce the original bug and pass after applying the corresponding gold patch, and the regression tests always pass. 
Table~\ref{tab:testcase} shows the test correctness of \tech{} across three benchmarks using two LLMs. 
Here, \textit{Base} denotes a straightforward LLM-based test generation approach that directly generates tests without additional guidance. 
We observe that \tech{} consistently and significantly outperforms \textit{Base}. 
In particular, the average test correctness of \tech{} ranges from 23.44\% to 46.41\%, whereas \textit{Base} achieves only 12.97\%. 
Furthermore, the correctness of tests generated by \tech{} improves as the number of iterations ($N$) increases. 
For example, \tech{}$_{N=4}$ improves test correctness by an average of 22.98\% compared to \tech{}$_{N=1}$. 
This trend suggests that iterative refinement enables \tech{} to progressively enhance test quality, and further gains may be achieved with larger $N$.

We acknowledge that a non-negligible proportion of tests remains imperfect, which is a common limitation across LLM-based test generation approaches and largely stems from the inherent constraints of current models. 
Nevertheless, both prior test-driven techniques~\cite{chen2025revisit, lei2025planning} and our empirical results demonstrate that even imperfect tests can substantially improve the overall effectiveness of software engineering tasks. 
In future work, we plan to explore more advanced test generation strategies (e.g., incorporating type-checking mechanisms and test selection techniques) to further improve test quality and patch generation performance of \tech{}.

\subsection{Orthogonality of \tech{} with Baselines}
\label{subsec:orthogonality}


In fact, \tech{} is largely orthogonal to existing issue resolution techniques. 
While \tech{} focuses on improving the quality of the original issue by constructing structured and information-rich issue-oriented requirements, existing techniques primarily optimize downstream stages, such as patch generation and validation, based on the given issue description.
As such, \tech{} can be viewed as a complementary pre-processing module that generates high-quality issue-oriented requirements before they are provided to existing techniques, thereby providing LLM-based agents with clearer and more precise guidance for issue resolution.
In future work, we plan to systematically investigate the integration of \tech{} with existing techniques to further enhance their overall performance.

\subsection{Future Work}
\label{subsec:future_work}

Although \tech{} has demonstrated strong effectiveness through comprehensive evaluation, several aspects can be further improved:

\begin{itemize}[leftmargin=10pt]
    \item \textbf{Improving Test Generation:}
    We plan to further enhance test generation by incorporating coverage-guided test generation strategies, type-checking mechanisms, and test selection approaches. 
    These improvements are expected to increase test quality, thereby enabling more accurate requirement refinement and more effective patch generation.
    
    \item \textbf{Improving Efficiency:}
    To reduce computational overhead, we aim to design more efficient context management strategies that compress historical context without discarding the essential information to preserve performance. 
    This would improve the practicality of \tech{} in real-world deployment scenarios.

    \item \textbf{Adaptive Iteration Strategies:}
    Currently, \tech{} employs a fixed number of iterations, which may lead to either insufficient exploration or unnecessary computational cost. 
    Future work will investigate adaptive iteration strategies that dynamically adjust the number of iterations based on feedback, enabling more efficient exploration of the solution space and further improving requirement quality and patch generation performance.
\end{itemize}

\section{Threats and Validity}
\label{sec:threats}



The threat to \textit{construct} validity primarily stems from the inherent randomness of LLM-based experiments. 
To mitigate this, we standardize key experimental settings across all techniques, including temperature and the maximum number of iterations, and conduct all experiments within a unified environment. 
Furthermore, we observe consistent performance trends across different settings (2 LLMs $\times$ 3 benchmarks), which increases confidence in the reliability of our results.
The threats to \textit{external} validity concern the generalizability of our findings. 
To alleviate this concern, we conduct evaluations on three widely used issue resolution benchmarks, employ two representative LLMs, and compare against five baselines using multiple evaluation metrics. 
Nevertheless, our experimental setup cannot fully capture the diversity of real-world software engineering scenarios. 
In future work, we plan to extend our evaluation to a broader range of benchmarks and LLMs.

\section{Related Work} 
\label{sec:related}

In recent years, the rapid advancement of LLMs has enabled the development of agents capable of generating code patches from issue descriptions to resolve repository-level software issues. 
According to prior studies~\cite{jiang2025agentic}, these approaches can be broadly categorized into agent-based and workflow-based issue resolution techniques.


\textbf{Agent-based} techniques equip LLM agents with tools for code navigation, editing, and execution, allowing them to autonomously explore the environment and iteratively improve patch generation. 
Representative techniques include academic systems such as SWE-agent~\cite{yang2024swe}, OpenHands~\cite{wangopenhands}, and AutoCodeRover~\cite{zhang2024autocoderover}, as well as industrial command-line interface (CLI) agents such as Claude Code~\cite{anthropic_claude_code}, Aider~\cite{aider2026}, Codex CLI~\cite{openai2025codexcli}, and Trae-agent~\cite{gao2025trae}. 
These approaches typically rely on interactive decision-making mechanisms that enable dynamic adaptation to environmental feedback, thereby supporting end-to-end issue resolution in realistic development settings.
In contrast, \textbf{workflow-based} techniques decompose the issue resolution process into a sequence of predefined stages, such as localization, patch generation, and patch validation. 
Representative techniques include PatchPilot~\cite{li2025patchpilot}, RepoGraph~\cite{ouyang2025repograph}, and Agentless~\cite{xia2025demystifying}. 
By explicitly modeling and controlling each stage, these approaches improve stability and reproducibility through structured execution pipelines.


Differing from these paradigms, \tech{} introduces a requirement-driven perspective that focuses on improving the foundational quality of issue specifications. 
By generating structured, information-rich issue-oriented requirements to effectively guide downstream patch generation, \tech{} is methodologically orthogonal and complementary to both agent-based and workflow-based techniques (as discussed in Section~\ref{subsec:orthogonality}).


\section{Conclusion}
\label{sec:conclusion}
In this work, we identify the quality of task inputs as a key bottleneck in repository-level issue resolution and introduce \textit{issue-oriented requirements} as structured task specifications to more effectively guide patch generation.
Building on this perspective, we propose \tech{}, a novel requirement-driven LLM agent framework that constructs structured and information-rich issue-oriented requirements, identifies low-quality requirements, and iteratively refines them to improve patch correctness.
To evaluate its effectiveness, we conduct comprehensive experiments on three widely used benchmarks using two advanced LLMs, comparing against five state-of-the-art and representative baselines. 
The results show that \tech{} consistently outperforms all baselines across multiple evaluation metrics, demonstrating its effectiveness for issue resolution tasks.


\begin{acks}

\end{acks}

\balance
\bibliographystyle{ACM-Reference-Format}
\bibliography{main}

@article{guo2024deepseek,
  title={DeepSeek-Coder: when the large language model meets programming--the rise of code intelligence},
  author={Guo, Daya and Zhu, Qihao and Yang, Dejian and Xie, Zhenda and Dong, Kai and Zhang, Wentao and Chen, Guanting and Bi, Xiao and Wu, Yifan and Li, YK and others},
  journal={arXiv preprint arXiv:2401.14196},
  year={2024}
}

@article{cao2026qwen3,
  title={Qwen3-Coder-Next Technical Report},
  author={Cao, Ruisheng and Chen, Mouxiang and Chen, Jiawei and Cui, Zeyu and Feng, Yunlong and Hui, Binyuan and Jing, Yuheng and Li, Kaixin and Li, Mingze and Lin, Junyang and others},
  journal={arXiv preprint arXiv:2603.00729},
  year={2026}
}

@inproceedings{jimenez2024swe,
  title={SWE-BENCH: CAN LANGUAGE MODELS RESOLVE REAL-WORLD GITHUB ISSUES?},
  author={Jimenez, Carlos E and Yang, John and Wettig, Alexander and Yao, Shunyu and Pei, Kexin and Press, Ofir and Narasimhan, Karthik},
  booktitle={12th International Conference on Learning Representations, ICLR 2024},
  year={2024}
}

@article{gao2025trae,
  title={Trae agent: An llm-based agent for software engineering with test-time scaling},
  author={Gao, Pengfei and Tian, Zhao and Meng, Xiangxin and Wang, Xinchen and Hu, Ruida and Xiao, Yuanan and Liu, Yizhou and Zhang, Zhao and Chen, Junjie and Gao, Cuiyun and others},
  journal={arXiv preprint arXiv:2507.23370},
  year={2025}
}

@inproceedings{han2024archcode,
  title={Archcode: Incorporating software requirements in code generation with large language models},
  author={Han, Hojae and Kim, Jaejin and Yoo, Jaeseok and Lee, Youngwon and Hwang, Seung-won},
  booktitle={Proceedings of the 62nd Annual Meeting of the Association for Computational Linguistics (Volume 1: Long Papers)},
  pages={13520--13552},
  year={2024}
}

@article{tian2025aligning,
  title={Aligning Requirement for Large Language Model's Code Generation},
  author={Tian, Zhao and Chen, Junjie},
  journal={arXiv preprint arXiv:2509.01313},
  year={2025}
}

@article{xia2025demystifying,
  title={Demystifying llm-based software engineering agents},
  author={Xia, Chunqiu Steven and Deng, Yinlin and Dunn, Soren and Zhang, Lingming},
  journal={Proceedings of the ACM on Software Engineering},
  volume={2},
  number={FSE},
  pages={801--824},
  year={2025},
  publisher={ACM New York, NY, USA}
}

@article{yang2024swe,
  title={Swe-agent: Agent-computer interfaces enable automated software engineering},
  author={Yang, John and Jimenez, Carlos E and Wettig, Alexander and Lieret, Kilian and Yao, Shunyu and Narasimhan, Karthik and Press, Ofir},
  journal={Advances in Neural Information Processing Systems},
  volume={37},
  pages={50528--50652},
  year={2024}
}

@article{deng2025swe,
  title={Swe-bench pro: Can ai agents solve long-horizon software engineering tasks?},
  author={Deng, Xiang and Da, Jeff and Pan, Edwin and He, Yannis Yiming and Ide, Charles and Garg, Kanak and Lauffer, Niklas and Park, Andrew and Pasari, Nitin and Rane, Chetan and others},
  journal={arXiv preprint arXiv:2509.16941},
  year={2025}
}

@article{yang2025qwen3,
  title={Qwen3 technical report},
  author={Yang, An and Li, Anfeng and Yang, Baosong and Zhang, Beichen and Hui, Binyuan and Zheng, Bo and Yu, Bowen and Gao, Chang and Huang, Chengen and Lv, Chenxu and others},
  journal={arXiv preprint arXiv:2505.09388},
  year={2025}
}

@article{bai2023qwen,
  title={Qwen technical report},
  author={Bai, Jinze and Bai, Shuai and Chu, Yunfei and Cui, Zeyu and Dang, Kai and Deng, Xiaodong and Fan, Yang and Ge, Wenbin and Han, Yu and Huang, Fei and others},
  journal={arXiv preprint arXiv:2309.16609},
  year={2023}
}

@inproceedings{wangopenhands,
  title={OpenHands: An Open Platform for AI Software Developers as Generalist Agents},
  author={Wang, Xingyao and Li, Boxuan and Song, Yufan and Xu, Frank F and Tang, Xiangru and Zhuge, Mingchen and Pan, Jiayi and Song, Yueqi and Li, Bowen and Singh, Jaskirat and others},
  booktitle={The Thirteenth International Conference on Learning Representations}
}

@inproceedings{zhang2024autocoderover,
  title={Autocoderover: Autonomous program improvement},
  author={Zhang, Yuntong and Ruan, Haifeng and Fan, Zhiyu and Roychoudhury, Abhik},
  booktitle={Proceedings of the 33rd ACM SIGSOFT International Symposium on Software Testing and Analysis},
  pages={1592--1604},
  year={2024}
}

@inproceedings{ruan2025specrover,
  title={Specrover: Code intent extraction via llms},
  author={Ruan, Haifeng and Zhang, Yuntong and Roychoudhury, Abhik},
  booktitle={2025 IEEE/ACM 47th International Conference on Software Engineering (ICSE)},
  pages={963--974},
  year={2025},
  organization={IEEE}
}

@article{li2025patchpilot,
  title={Patchpilot: A stable and cost-efficient agentic patching framework},
  author={Li, Hongwei and Tang, Yuheng and Wang, Shiqi and Guo, Wenbo},
  journal={arXiv e-prints},
  pages={arXiv--2502},
  year={2025}
}

@inproceedings{ouyang2025repograph,
  title={REPOGRAPH: ENHANCING AI SOFTWARE ENGINEERING WITH REPOSITORY-LEVEL CODE GRAPH},
  author={Ouyang, Siru and Yu, Wenhao and Ma, Kaixin and Xiao, Zilin and Zhang, Zhihan and Jia, Mengzhao and Han, Jiawei and Zhang, Hongming and Yu, Dong},
  booktitle={13th International Conference on Learning Representations, ICLR 2025},
  pages={30361--30384},
  year={2025},
  organization={International Conference on Learning Representations, ICLR}
}

@article{jiang2025agentic,
  title={Agentic Software Issue Resolution with Large Language Models: A Survey},
  author={Jiang, Zhonghao and Lo, David and Liu, Zhongxin},
  journal={arXiv preprint arXiv:2512.22256},
  year={2025}
}

@article{zhang2023survey,
  title={A survey on large language models for software engineering},
  author={Zhang, Quanjun and Fang, Chunrong and Xie, Yang and Zhang, Yaxin and Yang, Yun and Sun, Weisong and Yu, Shengcheng and Chen, Zhenyu},
  journal={arXiv preprint arXiv:2312.15223},
  year={2023}
}

@article{tao2024magis,
  title={Magis: Llm-based multi-agent framework for github issue resolution},
  author={Tao, Wei and Zhou, Yucheng and Wang, Yanlin and Zhang, Wenqiang and Zhang, Hongyu and Cheng, Yu},
  journal={Advances in Neural Information Processing Systems},
  volume={37},
  pages={51963--51993},
  year={2024}
}

@article{peng2023impact,
  title={The impact of ai on developer productivity: Evidence from github copilot},
  author={Peng, Sida and Kalliamvakou, Eirini and Cihon, Peter and Demirer, Mert},
  journal={arXiv preprint arXiv:2302.06590},
  year={2023}
}

@article{liu2023your,
  title={Is your code generated by chatgpt really correct? rigorous evaluation of large language models for code generation},
  author={Liu, Jiawei and Xia, Chunqiu Steven and Wang, Yuyao and Zhang, Lingming},
  journal={Advances in neural information processing systems},
  volume={36},
  pages={21558--21572},
  year={2023}
}

@inproceedings{xia2023automated,
  title={Automated program repair in the era of large pre-trained language models},
  author={Xia, Chunqiu Steven and Wei, Yuxiang and Zhang, Lingming},
  booktitle={2023 IEEE/ACM 45th International Conference on Software Engineering (ICSE)},
  pages={1482--1494},
  year={2023},
  organization={IEEE}
}

@article{shrivastava2023repofusion,
  title={Repofusion: Training code models to understand your repository},
  author={Shrivastava, Disha and Kocetkov, Denis and De Vries, Harm and Bahdanau, Dzmitry and Scholak, Torsten},
  journal={arXiv preprint arXiv:2306.10998},
  year={2023}
}

@article{jiang2026survey,
  title={A survey on large language models for code generation},
  author={Jiang, Juyong and Wang, Fan and Shen, Jiasi and Kim, Sungju and Kim, Sunghun},
  journal={ACM Transactions on Software Engineering and Methodology},
  volume={35},
  number={2},
  pages={1--72},
  year={2026},
  publisher={ACM New York, NY}
}

@article{meng2024empirical,
  title={An empirical study on llm-based agents for automated bug fixing},
  author={Meng, Xiangxin and Ma, Zexiong and Gao, Pengfei and Peng, Chao},
  journal={arXiv preprint arXiv:2411.10213},
  year={2024}
}

@inproceedings{yang2023users,
  title={What do users ask in open-source AI repositories? An empirical study of GitHub issues},
  author={Yang, Zhou and Wang, Chenyu and Shi, Jieke and Hoang, Thong and Kochhar, Pavneet and Lu, Qinghua and Xing, Zhenchang and Lo, David},
  booktitle={2023 IEEE/ACM 20th International Conference on Mining Software Repositories (MSR)},
  pages={79--91},
  year={2023},
  organization={IEEE}
}

@inproceedings{bettenburg2008makes,
  title={What makes a good bug report?},
  author={Bettenburg, Nicolas and Just, Sascha and Schr{\"o}ter, Adrian and Weiss, Cathrin and Premraj, Rahul and Zimmermann, Thomas},
  booktitle={Proceedings of the 16th ACM SIGSOFT International Symposium on Foundations of software engineering},
  pages={308--318},
  year={2008}
}

@article{zimmermann2010makes,
  title={What makes a good bug report?},
  author={Zimmermann, Thomas and Premraj, Rahul and Bettenburg, Nicolas and Just, Sascha and Schroter, Adrian and Weiss, Cathrin},
  journal={IEEE Transactions on Software Engineering},
  volume={36},
  number={5},
  pages={618--643},
  year={2010},
  publisher={IEEE}
}

@inproceedings{chaparro2017detecting,
  title={Detecting missing information in bug descriptions},
  author={Chaparro, Oscar and Lu, Jing and Zampetti, Fiorella and Moreno, Laura and Di Penta, Massimiliano and Marcus, Andrian and Bavota, Gabriele and Ng, Vincent},
  booktitle={Proceedings of the 2017 11th joint meeting on foundations of software engineering},
  pages={396--407},
  year={2017}
}

@inproceedings{davies2014s,
  title={What's in a bug report?},
  author={Davies, Steven and Roper, Marc},
  booktitle={Proceedings of the 8th ACM/IEEE International Symposium on Empirical Software Engineering and Measurement},
  pages={1--10},
  year={2014}
}

@article{soltani2020significance,
  title={The significance of bug report elements},
  author={Soltani, Mozhan and Hermans, Felienne and B{\"a}ck, Thomas},
  journal={Empirical Software Engineering},
  volume={25},
  number={6},
  pages={5255--5294},
  year={2020},
  publisher={Springer}
}

@article{liu2025deepseek,
  title={Deepseek-v3. 2: Pushing the frontier of open large language models},
  author={Liu, Aixin and Mei, Aoxue and Lin, Bangcai and Xue, Bing and Wang, Bingxuan and Xu, Bingzheng and Wu, Bochao and Zhang, Bowei and Lin, Chaofan and Dong, Chen and others},
  journal={arXiv preprint arXiv:2512.02556},
  year={2025}
}

@inproceedings{jainlivecodebench,
  title={LiveCodeBench: Holistic and Contamination Free Evaluation of Large Language Models for Code},
  author={Jain, Naman and Han, King and Gu, Alex and Li, Wen-Ding and Yan, Fanjia and Zhang, Tianjun and Wang, Sida and Solar-Lezama, Armando and Sen, Koushik and Stoica, Ion},
  booktitle={The Thirteenth International Conference on Learning Representations}
}

@article{chen2024coder,
  title={Coder: Issue resolving with multi-agent and task graphs},
  author={Chen, Dong and Lin, Shaoxin and Zeng, Muhan and Zan, Daoguang and Wang, Jian-Gang and Cheshkov, Anton and Sun, Jun and Yu, Hao and Dong, Guoliang and Aliev, Artem and others},
  journal={arXiv preprint arXiv:2406.01304},
  year={2024}
}

@article{pabba2025semagent,
  title={Semagent: A semantics aware program repair agent},
  author={Pabba, Anvith and Mathai, Alex and Chakraborty, Anindya and Ray, Baishakhi},
  journal={arXiv preprint arXiv:2506.16650},
  year={2025}
}

@article{chen2025prometheus,
  title={Prometheus: Unified knowledge graphs for issue resolution in multilingual codebases},
  author={Chen, Zimin and Pan, Yue and Lu, Siyu and Xu, Jiayi and Goues, Claire Le and Monperrus, Martin and Ye, He},
  journal={arXiv preprint arXiv:2507.19942},
  year={2025}
}

@article{wang2026swe,
  title={SWE-Pruner: Self-Adaptive Context Pruning for Coding Agents},
  author={Wang, Yuhang and Shi, Yuling and Yang, Mo and Zhang, Rongrui and He, Shilin and Lian, Heng and Chen, Yuting and Ye, Siyu and Cai, Kai and Gu, Xiaodong},
  journal={arXiv preprint arXiv:2601.16746},
  year={2026}
}

@inproceedings{lindenbauer2025complexity,
  title={The Complexity Trap: Simple Observation Masking Is as Efficient as LLM Summarization for Agent Context Management},
  author={Lindenbauer, Tobias and Slinko, Igor and Felder, Ludwig and Bogomolov, Egor and Zharov, Yaroslav},
  booktitle={NeurIPS 2025 Fourth Workshop on Deep Learning for Code}
}

@article{huang2019empirical,
  title={An empirical study on the issue reports with questions raised during the issue resolving process},
  author={Huang, Yonghui and da Costa, Daniel Alencar and Zhang, Feng and Zou, Ying},
  journal={Empirical Software Engineering},
  volume={24},
  number={2},
  pages={718--750},
  year={2019},
  publisher={Springer}
}

@inproceedings{ouhbi2013software,
  title={Software quality requirements: a systematic mapping study},
  author={Ouhbi, Sofia and Idri, Ali and Fern{\'a}ndez-Alem{\'a}n, Jose Luis and Toval, Ambrosio},
  booktitle={2013 20th Asia-Pacific Software Engineering Conference (APSEC)},
  volume={1},
  pages={231--238},
  year={2013},
  organization={IEEE}
}

@inproceedings{van2008requirements,
  title={Requirements engineering: from craft to discipline},
  author={Van Lamsweerde, Axel},
  booktitle={Proceedings of the 16th ACM SIGSOFT International Symposium on Foundations of software engineering},
  pages={238--249},
  year={2008}
}

@article{montgomery2022empirical,
  title={Empirical research on requirements quality: a systematic mapping study},
  author={Montgomery, Lloyd and Fucci, Davide and Bouraffa, Abir and Scholz, Lisa and Maalej, Walid},
  journal={Requirements Engineering},
  volume={27},
  number={2},
  pages={183--209},
  year={2022},
  publisher={Springer}
}

@article{stephen2020evaluation,
  title={Evaluation of software requirement specification based on IEEE 830 quality properties},
  author={Stephen, E and Mit, E},
  journal={International Journal on Advanced Science, Engineering and Information Technology},
  volume={10},
  number={4},
  pages={1396--1402},
  year={2020}
}

@article{suri2026codescout,
  title={CodeScout: Contextual Problem Statement Enhancement for Software Agents},
  author={Suri, Manan and Li, Xiangci and Shojaie, Mehdi and Han, Songyang and Hsu, Chao-Chun and Garg, Shweta and Deshmukh, Aniket Anand and Kumar, Varun},
  journal={arXiv preprint arXiv:2603.05744},
  year={2026}
}

@article{gervasi2002lightweight,
  title={Lightweight validation of natural language requirements},
  author={Gervasi, Vincenzo and Nuseibeh, Bashar},
  journal={Software: Practice and Experience},
  volume={32},
  number={2},
  pages={113--133},
  year={2002},
  publisher={Wiley Online Library}
}

@inproceedings{ferrari2014pragmatic,
  title={Pragmatic ambiguity detection in natural language requirements},
  author={Ferrari, Alessio and Lipari, Giuseppe and Gnesi, Stefania and Spagnolo, Giorgio O},
  booktitle={2014 IEEE 1st International Workshop on Artificial Intelligence for Requirements Engineering (AIRE)},
  pages={1--8},
  year={2014},
  organization={IEEE}
}

@article{aleithan2024swe,
  title={Swe-bench+: Enhanced coding benchmark for llms},
  author={Aleithan, Reem and Xue, Haoran and Mohajer, Mohammad Mahdi and Nnorom, Elijah and Uddin, Gias and Wang, Song},
  journal={arXiv preprint arXiv:2410.06992},
  year={2024}
}

@article{wang2025exploring,
  title={Exploring graph tasks with pure llms: A comprehensive benchmark and investigation},
  author={Wang, Yuxiang and Dai, Xinnan and Fan, Wenqi and Ma, Yao},
  journal={arXiv preprint arXiv:2502.18771},
  year={2025}
}

@article{mahran2025investigating,
  title={Investigating bias: A multilingual pipeline for generating, solving, and evaluating math problems with llms},
  author={Mahran, Mariam and Simbeck, Katharina},
  journal={arXiv preprint arXiv:2509.17701},
  year={2025}
}

@article{chen2026beyondswe,
  title={BeyondSWE: Can Current Code Agent Survive Beyond Single-Repo Bug Fixing?},
  author={Chen, Guoxin and Meng, Fanzhe and Zhao, Jiale and Li, Minghao and Cheng, Daixuan and Song, Huatong and Chen, Jie and Lin, Yuzhi and Chen, Hui and Zhao, Xin and others},
  journal={arXiv preprint arXiv:2603.03194},
  year={2026}
}

@article{rkaczkowska2023should,
  title={What should a good software requirements specification include? Results of a survey},
  author={R{\k{a}}czkowska-Gzowska, Karolina and Walkowiak-Gall, Anita},
  journal={Foundations of Computing and Decision Sciences},
  volume={48},
  number={1},
  pages={57--81},
  year={2023}
}

@inproceedings{pandey2010effective,
  title={An effective requirement engineering process model for software development and requirements management},
  author={Pandey, Dhirendra and Suman, Ugrasen and Ramani, A Kumar},
  booktitle={2010 International Conference on Advances in Recent Technologies in Communication and Computing},
  pages={287--291},
  year={2010},
  organization={IEEE}
}

@article{jin2024mare,
  title={Mare: Multi-agents collaboration framework for requirements engineering},
  author={Jin, Dongming and Jin, Zhi and Chen, Xiaohong and Wang, Chunhui},
  journal={arXiv preprint arXiv:2405.03256},
  year={2024}
}

@article{kruchten20024+,
  title={The 4+ 1 view model of architecture},
  author={Kruchten, Philippe B},
  journal={IEEE software},
  volume={12},
  number={6},
  pages={42--50},
  year={2002},
  publisher={IEEE}
}

@article{inkermann2019model,
  title={Model-based requirement engineering to support development of complex systems},
  author={Inkermann, David and Huth, T and Vietor, T and Grewe, A and Knieke, C and Rausch, A},
  journal={Procedia CIRP},
  volume={84},
  pages={239--244},
  year={2019},
  publisher={Elsevier}
}

@article{ramesh2021metrics,
  title={Metrics for software requirements specification quality quantification},
  author={Ramesh, MR Raja and Reddy, Ch Satyananda},
  journal={Computers \& Electrical Engineering},
  volume={96},
  pages={107445},
  year={2021},
  publisher={Elsevier}
}

@article{freund2012ieee,
  title={Ieee standard for system and software verification and validation (ieee std 1012-2012)},
  author={Freund, Eva},
  journal={Software quality professional},
  volume={15},
  number={1},
  pages={43},
  year={2012},
  publisher={American Society for Quality}
}

@article{terry2005requirements,
  title={Requirements development, verification, and validation exhibited in famous failures},
  author={Terry Bahill, A and Henderson, Steven J},
  journal={Systems engineering},
  volume={8},
  number={1},
  pages={1--14},
  year={2005},
  publisher={Wiley Online Library}
}

@article{umar2024advances,
  title={Advances in automated support for requirements engineering: a systematic literature review},
  author={Umar, Muhammad Aminu and Lano, Kevin},
  journal={Requirements Engineering},
  volume={29},
  number={2},
  pages={177--207},
  year={2024},
  publisher={Springer}
}

@article{franch2023state,
  title={The state-of-practice in requirements specification: an extended interview study at 12 companies},
  author={Franch, Xavier and Palomares, Cristina and Quer, Carme and Chatzipetrou, Panagiota and Gorschek, Tony},
  journal={Requirements engineering},
  volume={28},
  number={3},
  pages={377--409},
  year={2023},
  publisher={Springer}
}

@article{habiba2024mature,
  title={How mature is requirements engineering for AI-based systems? A systematic mapping study on practices, challenges, and future research directions},
  author={Habiba, Umm-e- and Haug, Markus and Bogner, Justus and Wagner, Stefan},
  journal={Requirements Engineering},
  volume={29},
  number={4},
  pages={567--600},
  year={2024},
  publisher={Springer}
}

@book{geisberger2007model,
  title={A Model-Based Approach To Requirements Analysis},
  author={Geisberger, Eva and Gr{\"u}nbauer, Johannes and Sch{\"a}tz, Bernhard},
  year={2007},
  publisher={Internat. Begegnungs-und Forschungszentrum f{\"u}r Informatik}
}

@article{windisch2022approach,
  title={Approach for model-based requirements engineering for the planning of engineering generations in the agile development of mechatronic systems},
  author={Windisch, Emily and Mandel, Constantin and Rapp, Simon and Bursac, Nikola and Albers, Albert},
  journal={Procedia CIRP},
  volume={109},
  pages={550--555},
  year={2022},
  publisher={Elsevier}
}

@article{bjarnason2014challenges,
  title={Challenges and practices in aligning requirements with verification and validation: a case study of six companies},
  author={Bjarnason, Elizabeth and Runeson, Per and Borg, Markus and Unterkalmsteiner, Michael and Engstr{\"o}m, Emelie and Regnell, Bj{\"o}rn and Sabaliauskaite, Giedre and Loconsole, Annabella and Gorschek, Tony and Feldt, Robert},
  journal={Empirical software engineering},
  volume={19},
  number={6},
  pages={1809--1855},
  year={2014},
  publisher={Springer}
}

@incollection{gigante2015semantic,
  title={A semantic driven approach for requirements verification},
  author={Gigante, Gabriella and Gargiulo, Francesco and Ficco, Massimo},
  booktitle={Intelligent distributed computing VIII},
  pages={427--436},
  year={2015},
  publisher={Springer}
}

@article{mucha2024systematic,
  title={A systematic literature review of pre-requirements specification traceability},
  author={Mucha, Julia and Kaufmann, Andreas and Riehle, Dirk},
  journal={Requirements Engineering},
  volume={29},
  number={2},
  pages={119--141},
  year={2024},
  publisher={Springer}
}

@article{yoo2024building,
  title={Building traceability between functional requirements and component architecture elements in embedded software using structured features},
  author={Yoo, Insun and Park, Hyoseung and Lee, Seok-Won and Ryu, Ki-Yeol},
  journal={Applied Sciences},
  volume={14},
  number={23},
  pages={10796},
  year={2024},
  publisher={MDPI}
}

@article{doe2011recommended,
  title={Recommended practice for software requirements specifications (ieee)},
  author={Doe, John},
  journal={IEEE, New York},
  year={2011}
}

@article{garcia2025greedy,
  title={Greedy algorithms: a review and open problems},
  author={Garc{\'\i}a, Andrea},
  journal={Journal of Inequalities and Applications},
  volume={2025},
  number={1},
  pages={11},
  year={2025},
  publisher={Springer}
}

@misc{openai_swebench_verified,
  author       = {{OpenAI}},
  title        = {Introducing SWE-bench Verified},
  year         = {2024},
  howpublished = {\url{https://openai.com/index/introducing-swe-bench-verified/}},
}

@misc{openai,
  author       = {{OpenAI}},
  title        = {About OpenAI},
  year         = {2026},
  howpublished = {\url{https://openai.com/about/}},
}

@misc{scaleai,
  author       = {{Scale AI}},
  title        = {About Scale AI},
  year         = {2026},
  howpublished = {\url{https://scale.com/about}},
}

@article{robertson1994okapi,
  title={Okapi at TREC},
  author={Robertson, Stephen Edward and Walker, Steve and Jones, Susan and Hancock-Beaulieu, Micheline M and Gatford, Mike and others},
  year={1994},
  publisher={British Library Board}
}

@inproceedings{han2026tdflow,
  title={TDFlow: Agentic Workflows for Test Driven Development},
  author={Han, Kevin and Maddikayala, Siddharth and Knappe, Tim and Patel, Om and Liao, Austen and Farimani, Amir Barati},
  booktitle={Proceedings of the 19th Conference of the European Chapter of the Association for Computational Linguistics (Volume 1: Long Papers)},
  pages={1511--1527},
  year={2026}
}

@inproceedings{chen2025revisit,
  title={Revisit self-debugging with self-generated tests for code generation},
  author={Chen, Xiancai and Tao, Zhengwei and Zhang, Kechi and Zhou, Changzhi and Zhang, Xinyu and Gu, Wanli and He, Yuanpeng and Zhang, Mengdi and Cai, Xunliang and Zhao, Haiyan and others},
  booktitle={Proceedings of the 63rd Annual Meeting of the Association for Computational Linguistics (Volume 1: Long Papers)},
  pages={18003--18023},
  year={2025}
}

@inproceedings{lei2025planning,
  title={Planning-driven programming: A large language model programming workflow},
  author={Lei, Chao and Chang, Yanchuan and Lipovetzky, Nir and Ehinger, Krista A},
  booktitle={Proceedings of the 63rd Annual Meeting of the Association for Computational Linguistics (Volume 1: Long Papers)},
  pages={12647--12684},
  year={2025}
}

@article{1963Critical,
  title={Critical Values and Probability Levels for the Wilcoxon Rank Sum Test and the Wilcoxon Signed Rank Test},
  author={ Wilcoxon, F.  and  Katti, S. K.  and  Wilcox, R. A. },
  year={1963},
}

@misc{anthropic_claude_code,
  author       = {{Anthropic}},
  year         = {2025},
  title        = {Claude Code: AI-powered coding assistant for developers},
  howpublished = {\url{https://www.anthropic.com/claude-code}},
}

@misc{aider2026,
  author = {Paul Gauthier},
  title = {Aider},
  year = {2025},
  howpublished = {\url{https://github.com/paul-gauthier/aider}}
}

@misc{openai2025codexcli,
  author = {OpenAI},
  title = {Codex CLI},
  year = {2025},
  howpublished = {\url{https://developers.openai.com/codex/cli}}
}

@inproceedings{tang2015will,
  title={Will this bug-fixing change break regression testing?},
  author={Tang, Xinye and Wang, Song and Mao, Ke},
  booktitle={2015 ACM/IEEE International Symposium on Empirical Software Engineering and Measurement (ESEM)},
  pages={1--10},
  year={2015},
  organization={IEEE}
}

@inproceedings{tian2026agent,
  title     = {Agent-Based Ensemble Reasoning for Repository-Level Issue Resolution},
  author    = {Tian, Zhao and Gao, Pengfei and Chen, Junjie and Peng, Chao},
  booktitle = {Proceedings of the 48th IEEE/ACM International Conference on Software Engineering (ICSE 2026)},
  year      = {2026},
}

@article{shur2024growing,
  title={Growing a Tail: Increasing Output Diversity in Large Language Models},
  author={Shur-Ofry, Michal and Horowitz-Amsalem, Bar and Rahamim, Adir and Belinkov, Yonatan},
  journal={Available at SSRN 5017241},
  year={2024}
}

@inproceedings{zhang2021trading,
  title={Trading off diversity and quality in natural language generation},
  author={Zhang, Hugh and Duckworth, Daniel and Ippolito, Daphne and Neelakantan, Arvind},
  booktitle={Proceedings of the Workshop on Human Evaluation of NLP Systems (HumEval)},
  pages={25--33},
  year={2021}
}

@inproceedings{minhturning,
  title={Turning Up the Heat: Min-p Sampling for Creative and Coherent LLM Outputs},
  author={Minh, Nguyen Nhat and Baker, Andrew and Neo, Clement and Roush, Allen G and Kirsch, Andreas and Shwartz-Ziv, Ravid},
  booktitle={The Thirteenth International Conference on Learning Representations}
}

@inproceedings{zanmulti,
  title={Multi-SWE-bench: A Multilingual Benchmark for Issue Resolving},
  author={Zan, Daoguang and Huang, Zhirong and Liu, Wei and Chen, Hanwu and Xin, Shulin and Zhang, Linhao and Liu, Qi and Li, Aoyan and Chen, Lu and Zhong, Xiaojian and others},
  booktitle={The Thirty-ninth Annual Conference on Neural Information Processing Systems Datasets and Benchmarks Track}
}

@article{kuang2025effectiveness,
  title={On the Effectiveness of Training Data Optimization for LLM-based Code Generation: An Empirical Study},
  author={Kuang, Shiqi and Tian, Zhao and Xiao, Tao and Wang, Dong and Chen, Junjie},
  journal={arXiv preprint arXiv:2512.24570},
  year={2025}
}

\end{document}